\documentclass[11pt,letterpaper]{article}
\usepackage[margin=1in]{geometry}

\usepackage{amsmath,amssymb}

\usepackage{changepage}

\usepackage[utf8x]{inputenc}

\usepackage{textcomp,marvosym}

\usepackage{cite}

\usepackage{nameref,hyperref}

\usepackage{microtype}
\DisableLigatures[f]{encoding = *, family = * }

\usepackage[table]{xcolor}

\usepackage{array}

\usepackage{wrapfig}

\usepackage{float}

\usepackage{newunicodechar}
\usepackage{libertine}
\DeclareRobustCommand{\okina}{%
  \raisebox{\dimexpr\fontcharht\font`A-\height}{%
    \scalebox{0.8}{`}%
  }%
}
\newunicodechar{ʻ}{\okina}

\newcolumntype{+}{!{\vrule width 2pt}}

\newlength\savedwidth

\usepackage[aboveskip=1pt,labelfont=bf,labelsep=period,justification=raggedright,singlelinecheck=off]{caption}

\makeatletter
\renewcommand{\@biblabel}[1]{\quad#1.}
\makeatother

\usepackage{lastpage,fancyhdr,graphicx}
\usepackage{grffile}
\usepackage{epstopdf}

\newcommand{\ignore}[1]{}
\newcommand{\Hawaii}{Hawai{\okina}i }
\newcommand{\Kauai}{Kaua{\okina}i }

\begin{document}
\vspace*{0.2in}

\begin{flushleft}
{\Large
\textbf\newline{COVID-19 Heterogeneity in Islands Chain Environment} 
}
\newline
\\
Monique Chyba\textsuperscript{1},
Alice Koniges\textsuperscript{2},
Prateek Kunwar\textsuperscript{1},
Winnie Lau\textsuperscript{1},
Yuriy Mileyko\textsuperscript{1},
Alan Tong\textsuperscript{1}
\\
\bigskip
\textbf{1} Applied and Computational Epidemiological Studies (ACES), University of \Hawaii at Manoa Department of Mathematics, Honolulu, \Hawaii, United States
\\
\textbf{2} \Hawaii Data Science Institute, University of \Hawaii at Manoa, Honolulu, \Hawaii, United States
\\
\bigskip

%
%






\end{flushleft}
\section*{Abstract}
As 2021 dawns, the COVID-19 pandemic is still raging strongly as vaccines finally appear and hopes for a return to normalcy start to materialize. There is much to be learned from the pandemic's first year data that will likely remain applicable to future epidemics and possible pandemics.  With only minor variants in virus strain, countries across the globe have suffered roughly the same pandemic by first glance, yet few locations exhibit the same patterns of viral spread, growth, and control as the state of Hawai'i. In this paper, we examine the data and compare the COVID-19 spread statistics between the counties of \Hawaii as well as examine several locations with similar properties to \Hawaii. 



\section*{Introduction}
Significant local variations in the spread of COVID-19 have been established in heterogeneous environments. For example, Thomas, et al., compares nineteen different cities and counties in the US\cite{Thomas24180}. They found that small differences in network models for interdependence and social interaction as well as the  effects due to  uneven  population  distributions  can  lead  to  substantial differences in infection timing and severity, leading different areas in each city to have vastly different experiences of the pandemic. Similar patterns associated with heterogeneity have been made for entire nations, such as the work comparing the most affected cities in China\cite{Cheng}. These works are based on the premise that substantial heterogeneity in social relationships at various scales affect the viral spread. It is unclear, however, whether or not such heterogeneity is a critical factor for an island chain and such study is absent from the literature. This is of utmost importance due to islands' vulnerability to any pandemic, especially for native populations as demonstrated for example with the introduction of measles to the Pacific Islands in the 1800's\cite{Shulman2009}. Islands are smaller contained populations, and thus epidemiological models may require adjustments to properly apply them to disease containment strategies. 
Identifying if major local variations can be expected for an island chain in the spread of COVID-19 is crucial since it directly impacts the effectiveness of mitigation measures, vaccine distribution and health care management. We focus here on a specific island chain, the Hawaiian archipelago and take somewhat different approach by comparing the individual island differences and identifying countries exhibiting similar properties. Maui for instance behave much more similarly to Japan over the last three months than her neighbor Islands which was surprising to see. Our goal is to demonstrate that Islands in general, whether they belong to the same archipelago or not, respond differently to the pandemic and cannot be aggregated into one single class. 
\begin{wrapfigure}{r}{0.45\textwidth}
    \centering
    \includegraphics[width=0.8\linewidth]{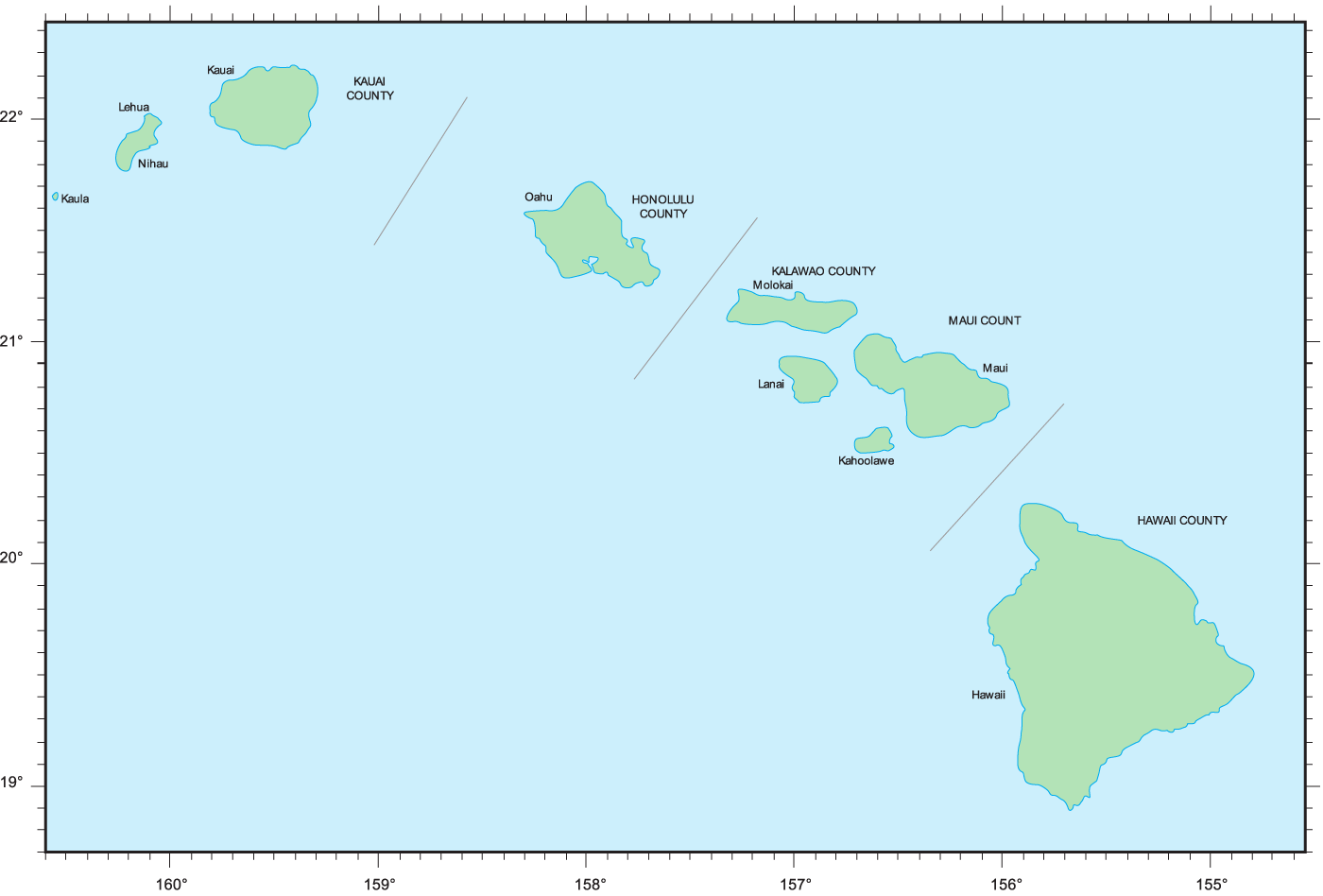}
    \caption{The State of \Hawaii and its counties\cite{counties-map}.}
    \label{fig:hawaii-map}
\end{wrapfigure}
The Hawaiian Islands are an archipelago of eight major islands, with only seven of them being populated. The State is divided into five counties: \Hawaii, Honolulu, Kalawao, \Kauai, and Maui.  Since Kalawao is the smallest county in all of the 50 states in terms of both population and land area, we focus here on only the four major countiesure \ref{fig:hawaii-map} shows the main eight islands as well as the various counties.\\

Table \ref{fig:state-general-statistics} shows that Honolulu city and county is the most populated county of the state, with 69\% of the state's population. \Hawaii county has the largest land mass of 63\% of the entire state, but comes second in resident population. Third by population is Maui county, which spans the islands of Maui, Moloka{\okina}i, Lanai, and Kaho{\okina}olawe. \Kauai county, which spans the islands \Kauai and Ni{\okina}ihau, has the smallest population. In determining heterogeneity effects and how the Hawaiian Islands might differ from each other, it is also important to compare the demographics of the four counties we study.
\begin{table}[h!]
    \centering
    \begin{tabular}{|c||c|c|c|c|}
        \hline
        Statistics &  Honolulu &  \Hawaii &  Maui &  \Kauai\\
        \hline
        \hline
        Land Area (miles) & 600.74 & 4,028.42 & 1,161.52 & 619.96\\
        Resident Population & 974,563 & 201,513 & 167,503 & 72,293\\
        Resident Population State Percent & 68.8\% & 14.2\% & 11.8\% & 5.1\%\\
        \hline
        Tourists (thousands per year) & 5862.4 & 1706.2 & 2914.9 & 1388.6\\
        Tourists (as daily percent of residents) & 1.64\% & 2.32\% & 4.77\% & 5.26\%\\
        \hline
    \end{tabular}
    \caption{The state's general statistics by county\cite{datawarehouse}.}
    \label{fig:state-general-statistics}
\end{table}

\begin{figure}[H]
    \centering
    \includegraphics[width=0.5\linewidth]{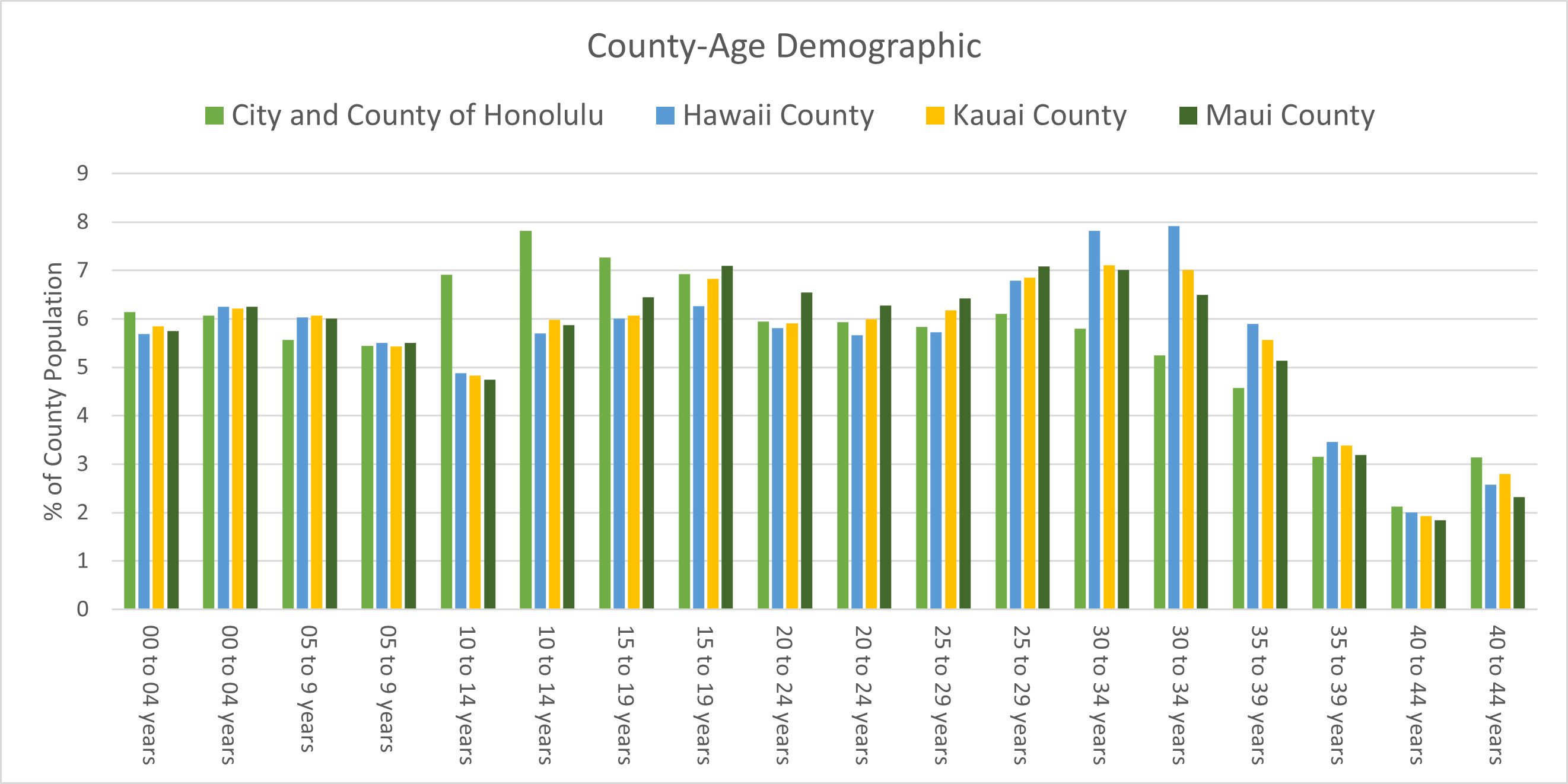}\includegraphics[width=0.5\linewidth]{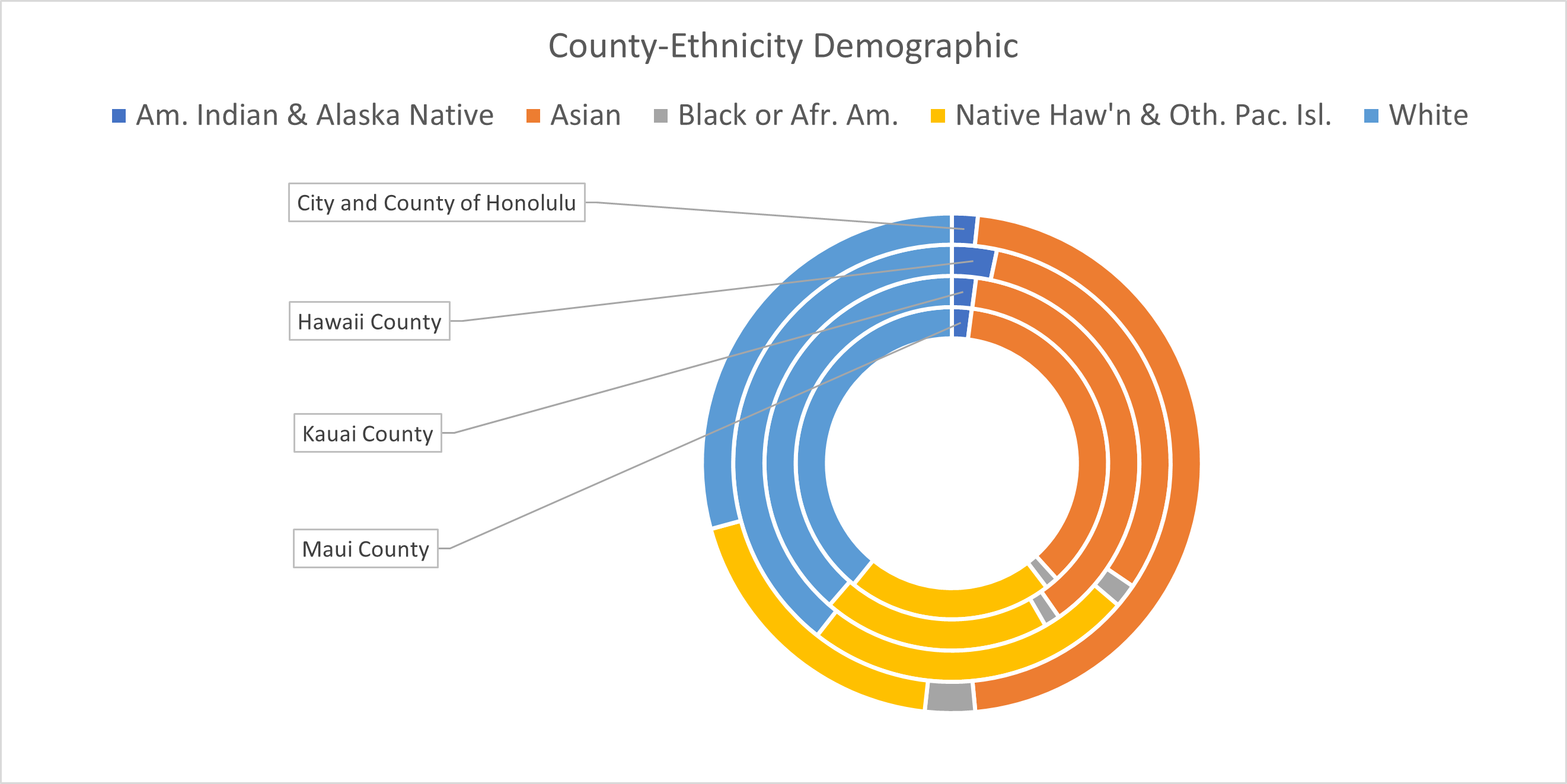}
    \caption{Left: Age demographic per county. Right: Ethnicity distribution per county\cite{HDC}.}
    \label{fig:agedemo}
\end{figure}

Figure \ref{fig:agedemo} left provides the age demographic distribution per county. Honolulu county has a larger percentage of individuals between 20 and 40 years old while \Hawaii county is more represented in the 55-80 years old age group. From Fig.~\ref{fig:agedemo} (right) it can be observed that the Honolulu county has a larger relative Asian population compared to the other counties and a smaller relative Native Hawaiian and Pacific Islander population.  

\begin{figure}[H]
    \centering
    \includegraphics[width=0.8\linewidth]{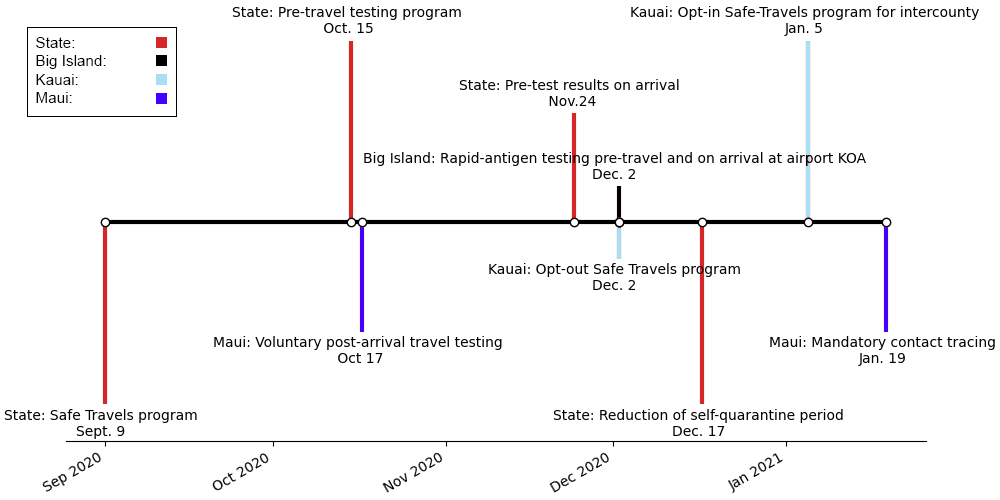}
    \caption{Safe travel protocols per counties. \Kauai county has the most restricted travel regulations since Dec. 2, 2020 following a significant initial surge in cases with the introduction of the Safe Travel program on October 15, 2020.}
    \label{fig:timeline travelers}
\end{figure}

While Honolulu city and county has been dominating the COVID-19 daily cases numbers due to its larger population, the other counties are also facing the pandemic. Intuitively we might expect all counties to exhibit homogeneity with respect of impact of the virus, however this is not observed. We describe in detail commonalities and differences between the four counties. Additionally,  we compare them to other non-Hawaiian islands to find similarities and differences. Our work highlights the need for localized measures and possibly targeted mitigation measures at the county level and as opposed to the state level for the most effective pandemic control. This has been already initiated to some degree with \Kauai county implementing their more restricted travel policy on Dec. 2, 2020; on Jan. 19, 2021, Maui implemented the mandatory safe travel app for all travelers, see Fig. \ref{fig:timeline travelers} for more details. It is critical for decision makers to take into account heterogeneity in their strategies. 

An important conclusion of this research is the identification of patterns that change extremely rapidly. This is due primarily to the nonlinear behaviour of the underlying equations that simulate the spread of the pandemic. In other words, it is not sufficient to average the initial conditions of the virus spread and assume that the different islands will exhibit similar behavior in an average sense. On the contrary, nonlinear effects and clusters can take off in one of the contained populations at a different time, thus requiring different pandemic control mandates. We find that it is critical to assure that heterogeneity is included in modeling and thus decision making for adequate and effective pandemic control. 


\section*{Materials and methods}
There are useful collections of \Hawaii COVID-19 data in the form of dashboards: the \Hawaii Emergency Management Agency’s (HiEMA) dashboard, the State of \Hawaii’s Department of Health’s Disease Outbreak Control Division’s (DOCD) COVID-19 dashboard, and COVID Pau dashboard (CPD) \cite{HiEMA, DOH, Pau}. Directly utilizing these dashboards alone is challenging. Firstly, the dashboards are not synchronized; they often display different data at various times for the same quantities, such as hospitalization data. Secondly, the availability of the dashboard data is sometimes restricted because of political concerns. Both HiEMA and DOCD provide visual data in plots, but do not allow for downloading of the data. The \Hawaii Data Collaborative dashboard\cite{HDC} resolves a majority of these issues by providing a Google Spreadsheet of the local Department of Health's DOCD data. The \Hawaii Data Collaborative also works to coalesce data from the other dashboards, and even obtains data directly from the office of Lt. Governor Josh Green. Collected data and their sources are summarized in Table \ref{tab:data-source}.

\begin{table}[h!]
    \centering
        \begin{tabular}{|c|c|}
        \hline
        Statistic & Source\\
        \hline
        Daily Cases & \Hawaii Data Collaborative\cite{HDC}\\
        Deaths & \Hawaii Data Collaborative\cite{HDC}\\
        Testing Data & \Hawaii State Department of Health\cite{DOH}\\
        Hospitalization & \Hawaii Data Collaborative\cite{HDC}\\
        Infections by County & \Hawaii State Department of Health\cite{DOH}\\
        Mobility Index & \Hawaii State Department of Health\cite{DOH}\\
        Traveler Data & \Hawaii Data Collaborative\cite{HDC}\\
        \hline
        \end{tabular}
        \caption{The sources of COVID-19 statistics for this paper. The \Hawaii State Department of Health data is original, while the \Hawaii Data Collaborative takes a large portion of it's data from the Department of Health.}
    \label{tab:data-source}
\end{table}

We also use the distribution of cases per zip code for each county whose tabulation areas are illustrated in \nameref{S1_Fig-Zipcode}. This data proved even more challenging to gather, since those numbers are not compiled in any open source spreadsheet; they need to be fetched from the Disease Outbreak Control Division Dashboard under their \Hawaii COVID-19 Maps daily. To obtain mobility data we used the open source SafeGraph COVID-19 Data Consortium \cite{Safegraph} that provides social distancing metrics illustrating the daily view of movement between census block groups. 

\begin{figure}[H]
    \centering
    \includegraphics[width=1.0\linewidth]{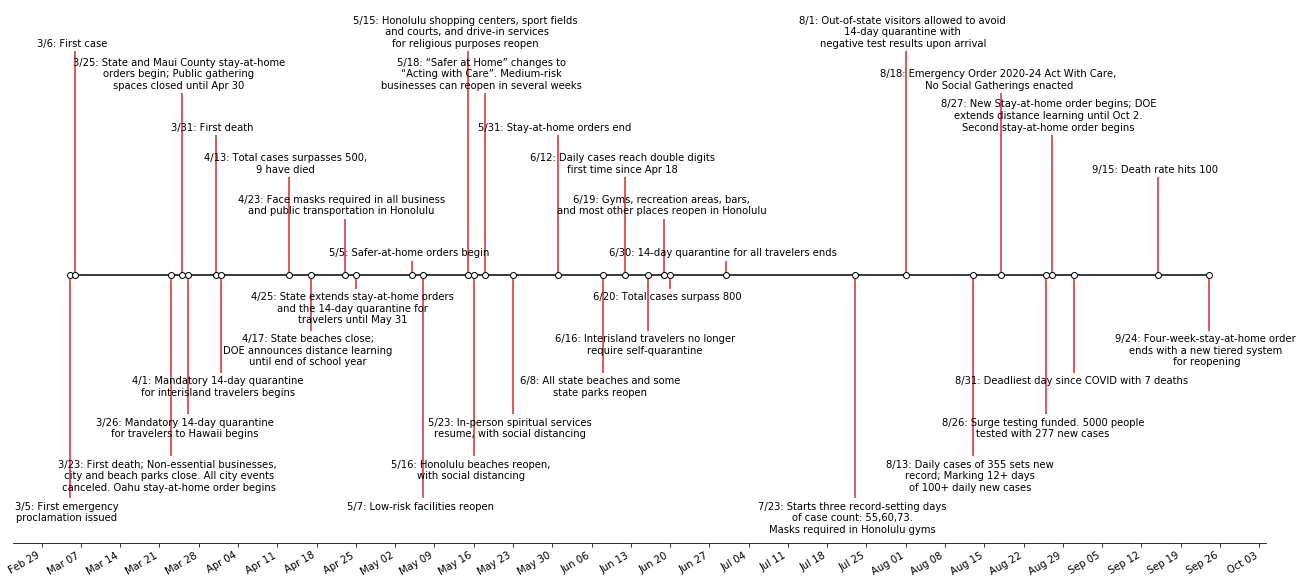}
    \caption{Timeline of events related to the pandemic in the State of \Hawaii from March 6, 2020 to September 24, 2020.}
    \label{fig:timeline State}
\end{figure}

The transmission rate in our model is optimized to reflect non pharmaceutical mitigation interventionsure \ref{fig:timeline State} displays the timeline from March 6 to September 24. Primary events impacting the curve after September 24 are due to the {\it safe travel program} and can be seen in Fig.~\ref{fig:timeline travelers}. In addition, the State moved to Tier 2 on October 22, 2020 and has stated in that phase since. Note also that the State of \Hawaii started vaccines administration on Dec 15, 2020. As of January 17, 2021 the State recorded 76'498 administered vaccines doses. The deadliest day since September 24, and global maximum happened on October 14, 2020 with a count of 14 individual. 
In this paper, we also compare the Hawaiian counties to other countries, Table \ref{tab:data-source-countries} summarizes the sources for the data we used. 
\begin{table}[h!]
    \centering
        \begin{tabular}{|c|c|}
        \hline
        Statistic & Source\\
        \hline
        Daily Cases Iceland & COVID-19 in Iceland - Statistics\cite{Icelanddata}\\
        Daily Cases Japan & Japan COVID-19 Coronavirus Tracker\cite{Japandata}\\
        Daily Cases Puerto-Rico & The COVID Tracking Project\cite{PRdata}\\
        \hline
        \end{tabular}
        \caption{The sources of COVID-19 data used in this paper for the comparison countries.}
    \label{tab:data-source-countries}
\end{table}

\subsection*{Compartmental Model}
There are two main classes of epidemiological models for this type of disease spread: compartmental models\cite{Brauer,Andrea,getz_dougherty2018} and agent-based models \cite{Covasim,Hoertel,perez_dragicevic2009,hunter_etal2018}. In this paper, we use a compartmentalized model inspired by \cite{Lloyd}, which is based on a standard discrete SEIR model. An extension, key to this paper, that we added to the model is a new group for travelers. Indeed, the tourist population plays a prominent role in \Hawaii and due to our isolated geographic location we are able to to collect precise information about daily arrivals and departure. 

In our model, a given population is divided into four compartments: Susceptible (not currently infected), Exposed (infected with no symptoms), Infected (infected with symptoms), and Removed (recovered or deceased). Moreover, we subdivide the entire population into three additional groups: the general community (C), healthcare workers (H) and visitors (V). Visitors, who are only considered after October 15, when the safe travels \Hawaii program began, are further broken down into two categories: returning residents and tourists. While the returning residents are absorbed into the community bucket, the tourists are treated as a separate group.  These groups interact with each other, and each of them consists of the aforementioned compartments. In addition, compartments Exposed and Infected (in each population group) are split into multiple stages by day to better reflect the progression of the disease. There are two key dynamics of each population group: the dynamics of Susceptible individuals and the dynamics of the rest of the compartments. The time dependent \emph{hazard rate}, $\lambda(t)$, governs the susceptible dynamics as it determines the probability, $1-e^{-\lambda(t)}$, of an individual becoming exposed at time $t$. The hazard rate is different for different population groups and takes into account interactions between the groups, thus coupling their dynamics. 

Key to governing the spread of the disease is parameter $\beta$, capturing the basal transmission rate due to various interactions among individuals. Our model optimizes $\beta$ to fit daily cases for a specific geographic location. Specifically, we use several different values of $\beta$ that capture changes in COVID-19 mitigation policy. Table \ref{tab:parameters-common} displays the variables and parameters common to all simulations in this paper (optimized $\beta$'s are given in the Results section). We introduce parameters $p_i$ as probabilities to develop symptoms on day $i$, and chose them such that if symptoms do develop, it takes between 2 to 14 days, with a mean between 4 and 6 days\cite{park2020systematic}, while assuming that about 40\% of all infections remain asymptomatic. The values of $q_{s,i}$ reflect the sentiment that symptomatic individuals are likely to quarantine, especially after a couple of days of symptoms. In addition, parameter $r$ is the probability of transitioning from one stage of the illness to the next (with the final stage being recovery or death). Based on prior work \cite{GlopbalHealth2}, we chose $r$ to yield an expected length of illness of 17 days. 
\begin{table}[h!]
\caption{Variable and parameters common for all geographic locations}
\centering
\begin{tabular}{ |>{\raggedright\let\newline\\\arraybackslash\hspace{0pt}}p{6.5cm}||>{\raggedright\let\newline\\\arraybackslash\hspace{0pt}}p{6cm}|}
 \hline
\centerline {Parameter, meaning} & \centerline {Value} \\
 \hline
  \hline
$\beta$, basal transmission rates   &  optimized to fit data  \\
 \hline
  \multicolumn{2}{|c|}{Factors modifying transmission rate}\\
 \hline
$\varepsilon$, asymptomatic transmission & 0.75 \\
 \hline
$\rho$, reduced healthcare worker interactions &  0.8 \\
  \hline
  $\rho_{v}$, reduced visitor-community interaction & 0.5 \\
  \hline
     $\gamma$, quarantine &  0.2   \\
   \hline
   $\gamma_{v}$, quarantine for visitor & 0.3 \\
  \hline
   $\kappa$, hospital precautions & 0.5\\
   \hline
   $\eta$, healthcare worker precautions & $0.2375$\\
   \hline
  \multicolumn{2}{|c|}{Population fractions}\\
   \hline
  $p_i$, $i=$0,\ldots,13, onset of symptoms after day $i$  & 0.000792, 0.00198, 0.1056, 0.198, 0.2376, 0.0858, 0.0528, 0.0462,
0.0396, 0.0264, 0.0198, 0.0198, 0.0198, 0\\
  \hline
  $q_{s,i}$, $i=$0,\ldots,4, symptomatic quarantine after day/stage $i$ & C: 0.1, 0.4, 0.8, 0.9, 0.99;\newline H: 0.2, 0.5, 0.9, 0.98, 0.99\\
  \hline
  r, transition to next symptomatic day/stage & 0.2\\
   \hline
   $\nu$, symptomatic hospitalization & 0.075\\
  \hline
  $\iota$, icu admission rate of hospitalized patients & 0.2 \\
  \hline
  
\hline
\end{tabular}
\label{tab:parameters-common}
\end{table}

In addition, we have parameters related to mitigation measures such as mask compliance as well as contact tracing that depend on the geographical location. Table \ref{tab:parameters} lists the values we use for the State of \Hawaii (those are assumed to be constant over the various counties) as well as the ones for others countries relevant to the discussion section. The parameters have been identified from dashboards/articles as well as for contact tracing. The choice of $q_{a,i}$ reflects the various testing and contact tracing efforts, and provide the probability for an asymptomatic individual to go into isolation as a result of testing and contact tracing.  
\begin{table}[h!]
\caption{Geographically dependent factors modifying transmission rate}
\centering
\begin{tabular}{ |>{\raggedright\let\newline\\\arraybackslash\hspace{0pt}}p{3.3cm}||>{\raggedright\let\newline\\\arraybackslash\hspace{0pt}}p{2cm}||>{\raggedright\let\newline\\\arraybackslash\hspace{0pt}}p{2cm}||>{\raggedright\let\newline\\\arraybackslash\hspace{0pt}}p{2cm}||>{\raggedright\let\newline\\\arraybackslash\hspace{0pt}}p{2cm}|}
 \hline
\centerline {Parameter, meaning} & \centerline {HI Counties} & \centerline {Japan} & \centerline{Puerto Rico} & \centerline{Iceland}\\
 \hline
  \hline
  \multicolumn{5}{|c|}{Factors modifying transmission rate}\\
 \hline
  \hline
   
   $p_{mp}$, mask compliance & 0.2 before Aug 27, 0.7 thereafter & 0.2 before May 04, 0.8 thereafter & 0.2 before Aug 21, 0.7 thereafter & 0.2 before Oct 20, 0.5 thereafter\\
   \hline
   $p_{me}$, mask efficiency & 0.25 & 0.25 & 0.25 & 0.25\\
   \hline
  \multicolumn{5}{|c|}{Population fractions}\\
  
  \hline
  $q_{a,i}$, $i=$0,\ldots,13, asymptomatic quarantine after day $i$ & 0 before Jun 08, then $q_5=q_6=q_7=$ 0.05  & 0 before Feb 25, then $q_5=q_6=q_7=$ 0.05& 0 before May 05, then $q_5=q_6=q_7=$ 0.05 & 0 before Apr 01, then $q_5=q_6=q_7=$ 0.05\\

\hline

\end{tabular}
\label{tab:parameters}
\end{table}

For more information regarding dynamics equations of the model, see \nameref{S1_Appendix_Model}. 
\section*{Results}
In this section we provide the results of simulations of our model for the four counties of the State of \Hawaii under analysis. In our plots we use the raw daily cases and not the 7 day average because our model fit plots the sum over all groups of the newly isolated and quarantined daily exposed and infected individuals, see S1 Appendix, Model Dynamics.
\paragraph{Initial Conditions.}
The initial values of most variables are zero. The only non-zero values are the number of susceptible individuals in the general community and the  healthcare worker community, the values of which are listed in Table \ref{table-susceptible-population-Hawaii}, as well as a single not quarantined symptomatic individual, $I_{c,0}(0)=1$. 
\begin{table}[h!]
    \centering
    \begin{tabular}{|c|c|c|c|}
        \hline
         Region &  $S_c(0)$ & $S_h(0)$ & Date for $I_{c,0}(0)=1$\\
        \hline
        \hline
        Honolulu & 937711 & 15000 & Mar 06 \\
        \hline
        Maui & 167417 & 1500 & Mar 15\\
        \hline
        \Hawaii & 201513 & 1500 & Mar 16\\
        \hline
    \end{tabular}
    \captionsetup{justification=centering}
    \caption {Susceptible population for each region and first detected symptomatic individual. All other variables have an initial value of 0.}
    \label{table-susceptible-population-Hawaii}
\end{table}
\subsection*{Honolulu County}
\begin{figure}[H]
    \centering
    \includegraphics[width=\linewidth]{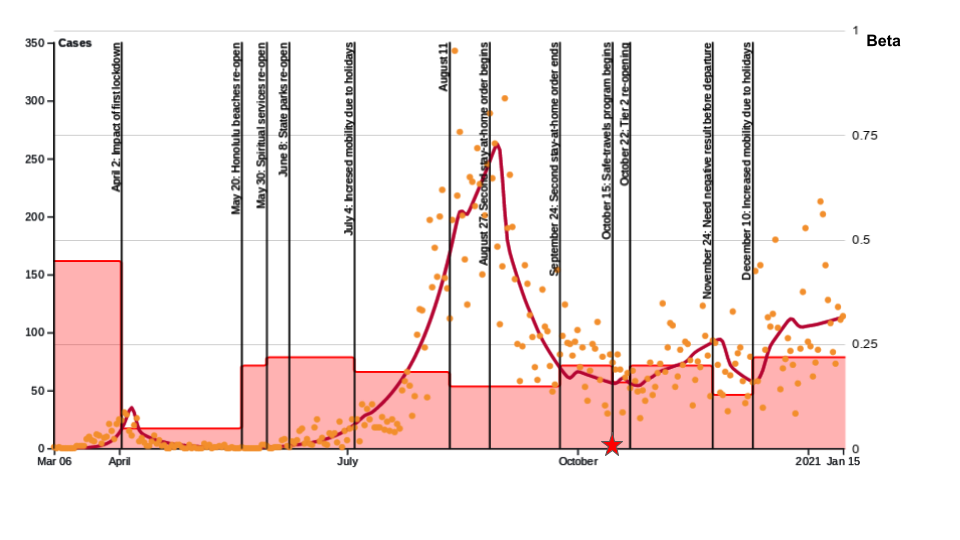}
    \caption{City and county of Honolulu: COVID-19 daily count and key events reflecting a change in behavior. The star shows the beginning of safe travel program. }
    \label{fig:honolulu-plot-fit}
\end{figure}

\begin{figure}[H]
    \centering
    \includegraphics[width=\linewidth]{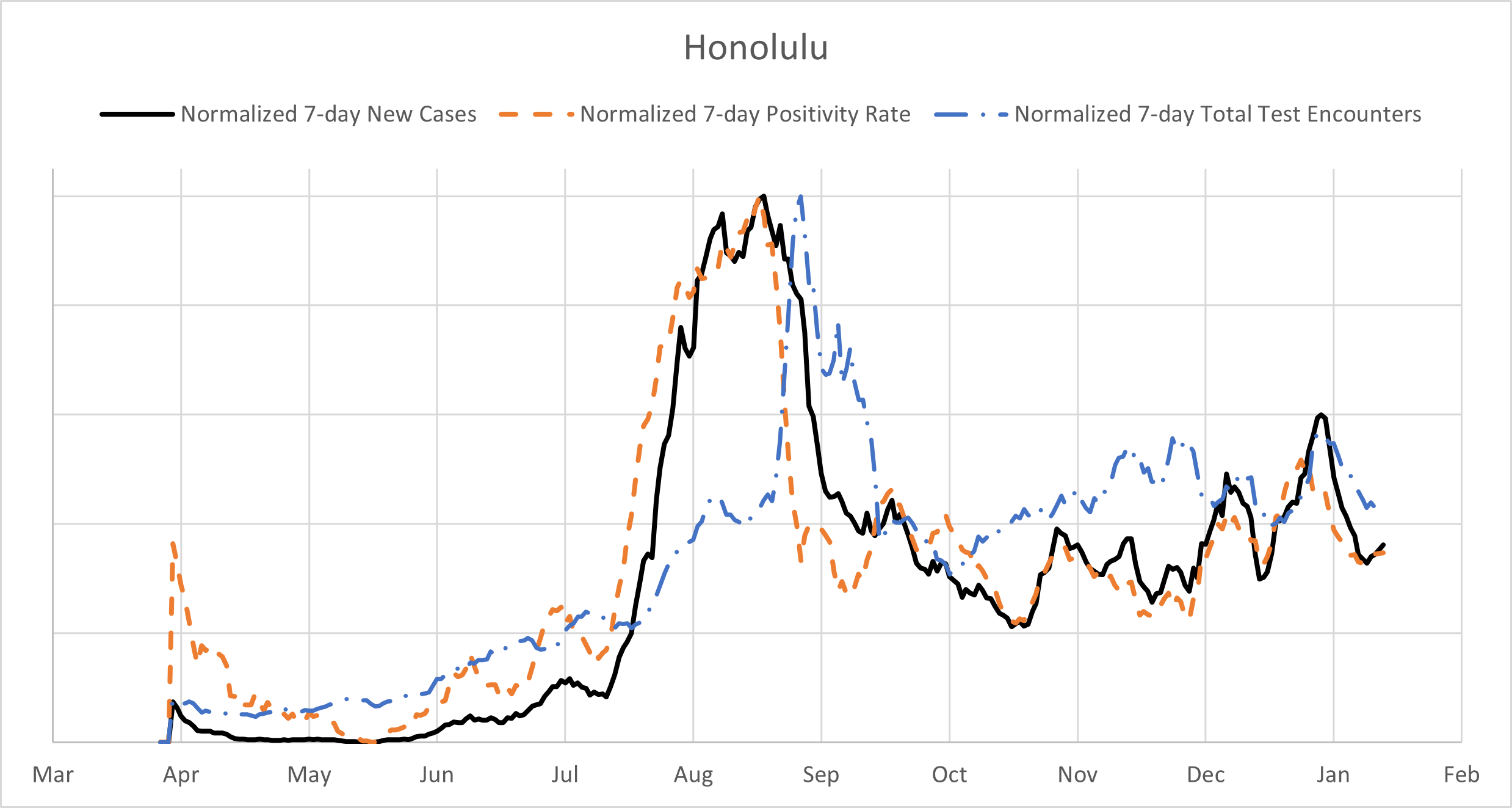}\\
    \includegraphics[width=\linewidth]{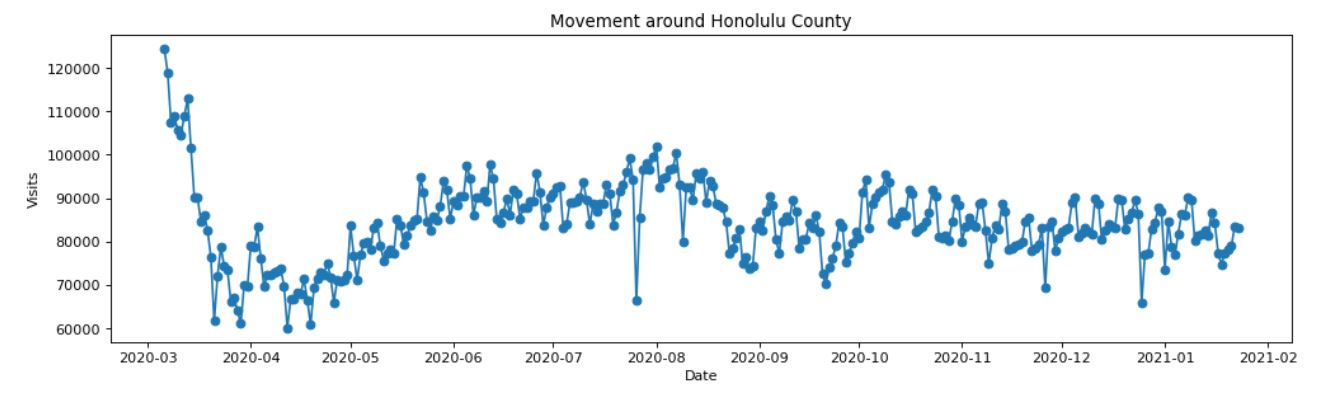}
    \caption{Top: A sharp increase in the test positivity rate (along with the daily cases) in July indicates an outbreak of the disease. The later decrease in the positivity rate with the increased number of tests indicates a substantial slowdown of the spread of the disease. Bottom: Overall mobility for Honolulu County from March 2020 to January 2021 suggest a modest correlation with the number of daily cases. }
    \label{fig:positivity-honolulu}
\end{figure}

Figure \ref{fig:honolulu-plot-fit} displays the model fit for the Honolulu county. The dots represents the daily cases and the curve is the model fit. The vertical lines corresponds to mitigation measures that had an impact on the curve and for which we optimized the $\beta$. Table \ref{Table-transmissionrate-Oahu} explicit the different $\beta$'s. The maximal daily case for Honolulu county was 342 and happened on August 12, 2020. We see two major exponential growths, one early in March that was crushed through a stay-at-home order and one in August followed by a second stay-at-home order. However the second lockdown was lifted before daily cases reached single digits in the hope to save the local economy. It can be seen on Table \ref{Table-transmissionrate-Oahu} that the first lockdown was more efficient. The largest peak is attributed to the July 4 festivities, the transmission rate $\beta$ was however quite smaller than for the first peak, but the State was much slower to call for a second stay-at-home order which resulted in the significantly higher counts. On October 15, 2020 the state of \Hawaii introduced the safe travel program which prompted an influx of tourists and traveling residents, this influx varies with time which explains the waving shape of the fit. For more details on incorporation of travelers in our model see \nameref{S1_Appendix_Model}. Since the Safe travel program the daily cases have been fluctuating quite a lot which makes a fit difficult (some high daily cases came from a correctional facility cluster for instance). The overall trend as of January 15, 2021 is shown to be slightly increasing (the 7-day average can be found in Fig.\ref{fig:positivity-honolulu}).   

\begin{figure}[H]
    \centering
    \includegraphics[width=\linewidth]{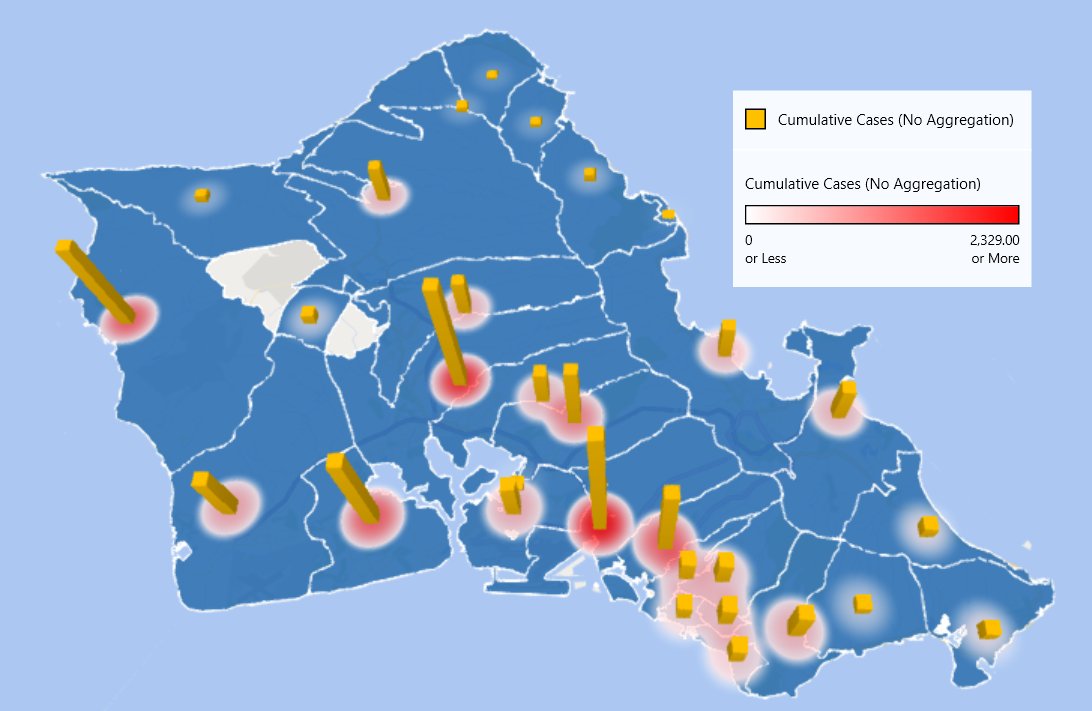}
    \caption{Honolulu county cumulative daily counts distributed per zip code from March 2020 to January 18, 2021.}
    \label{fig:honolulu3D-zip-code}
\end{figure}

\begin{figure}[H]
    \centering
    \includegraphics[width=0.5\linewidth]{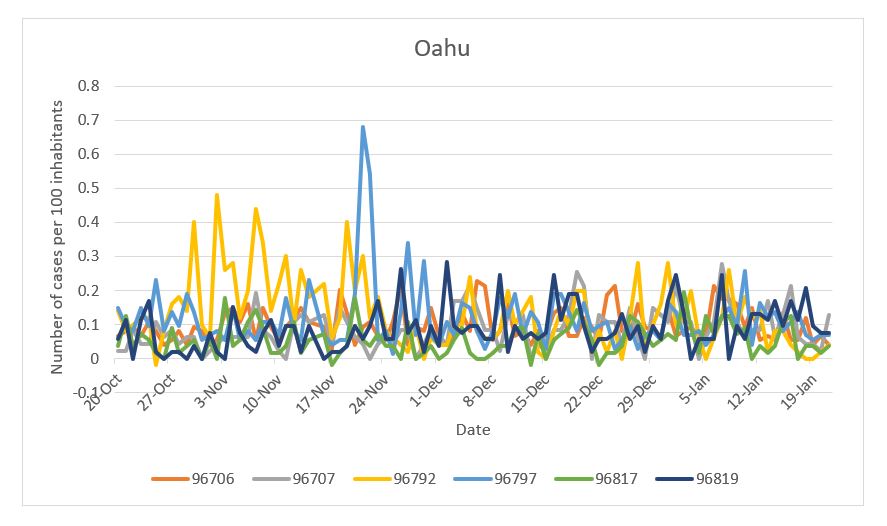}\includegraphics[width=0.5\linewidth]{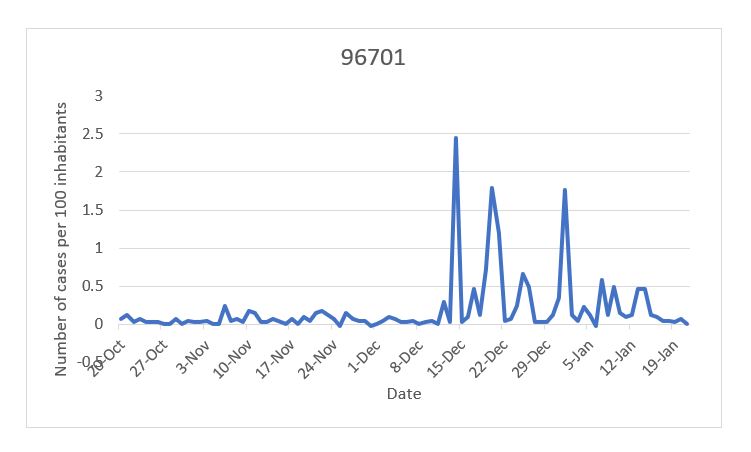}
    \caption{Honolulu county cumulative daily counts distributed per zip code from October 2020 to January 18, 2021.}
    \label{fig:honolulu-zip-code-plots}
\end{figure}

\begin{table}[h!]
    \centering
    \begin{tabular}{ |p{2.8cm}|p{2.8cm}|p{2.8cm}|p{2.8cm}|}
        \hline
        \multicolumn{4}{|c|}{Transmission rates} \\
        \hline
        March 6 - April 1 & April 2 - May 19 & May 20 - May 29 & May 30 - Jul 3\\
        \hline
        $\beta=0.45$ & $\beta=0.05$   & $\beta=0.2$ & $\beta=0.22$\\
        \hline
          Jul 4 - Aug 10 & Aug 11 - Sep 23 & Sep 24 - Oct 14 & Oct 15 - Oct 21\\
        \hline
          $\beta=0.185$ & $\beta=0.15$ & $\beta=0.2$ & $\beta=0.16$\\
        \hline
         Oct 22 - Nov 23 & Nov 24 - Dec 9& Dec 10 - Jan 15  & \\
        \hline
        $\beta=0.20$ & $\beta=0.13 $ & $\beta=0.22$ & \\
        \hline
    \end{tabular}
    \caption{Optimized transmission rates to fit Honolulu county data. They reflect the State and Honolulu non-pharmaceutical mitigation measures. }
    \label{Table-transmissionrate-Oahu}
\end{table}
The top of Fig.\ref{fig:positivity-honolulu} shows the total number of tests, the test positivity rate (i.e. the percent of tests for COVID-19 that came back positive) as well as the daily cases for the Honolulu County. To create this overlayed plot, the shown metrics have been normalized by calculating each data point as a percent of the maximum of the corresponding metric over the whole observation period and using the 7-day rolling average. It can be seen that, as anticipated, test positivity correlates strongly with daily cases. The noticeably large initial values of the test positivity rate (also present for other counties) are likely caused by the a small number of test that have been administered to a very narrow slice of the population with much higher chances of having the virus. When interpreting these plots, it should also be noted that even later in the pandemic the sample of people receiving tests was not unbiased, since the State of \Hawaii has been administering tests to people who satisfy criteria which make them more likely to have the virus. The bottom plot of Fig.\ref{fig:positivity-honolulu} displays the mobility for Honolulu County, it shows the major dip in mobility triggered by the first stay-at-home order back in March 2020, coming back up in May to peak again in August before the second stay-at-home order. The mobility data clearly suggests why the second lockdown was not as efficient as the first one.
In addition to the daily cases, we represents in Fig.\ref{fig:honolulu3D-zip-code} the cumulative daily counts for Honolulu county distributed per zip code from the onset of daily cases to January 18, 2021.  It can be observed that Honolulu downtown as well as the West Coast (Waianae) have been the most affected in terms of daily cases. For the West Coast it is mostly due to its high pacific islanders population and the fact that they have been disproportionately impacted. While they form about 4\% of the total \Hawaii population they account for more than 27\% of total cases \cite{samoa}. 
Table \ref{tab:cumulative-data-zip-code-honolulu} highlights the numbers for the seven highest zip codes. Zip code 96819 dominates the count per 100 inhabitants, containing Moanalua, Kalihi, Kapalama, and Daniel K. Inouye International Airport on the south side of Oahu. The second one is 96792 of the Waianae area on the west side of Oahu. From Fig.\ref{fig:honolulu-zip-code-plots}, we see that zip code 96701 displays a cluster behavior and that almost all its cases happened between December 16, 2020 to January 6, 2021. This was due to a cluster at Halawa Correctional Facility. There is no real immediate visible pattern from the other zip codes.
\begin{table}[h!]
    \centering
       \begin{tabular}{ |p{2.8cm}|p{2.8cm}|p{2.8cm}|p{2.8cm}|}
        \hline
        Honolulu County Zip code & Population Estimate & Cumulative Daily cases & Cum. Daily cases per 100 inhabitants\\
        \hline\hline
         96701 & 40857 & 1156 & 28 \\
         96706 & 74592 & 1562 & 21 \\
         96707 & 46928 & 850 & 18 \\
         96792 & 49971 & 1534 & 31 \\
         96797 & 73579 & 2038 & 28 \\
         96817 & 56144 & 1493 & 26.5 \\
         96819 & 52981 & 2342 & 44 \\
        \hline
    \end{tabular}
    \caption{The seven zip codes with the largest cumulative distribution of daily cases.}
    \label{tab:cumulative-data-zip-code-honolulu}
\end{table}

\subsection*{\Hawaii County}
Daily cases for \Hawaii county were very small until the aftermath of the July 4 celebrations which generated a large spike. The second stay-at-home order on Maui was extremely efficient but immediately followed by an exponential increase in the form of a few clusters. The maximum value is 51 and happened on October 25, 2020 during the third peak with a very close value during the second peak of 39 on August 29, 2020. One can observe a somewhat puzzling decrease in the number of daily cases after the start of the safe travel program. A potential explanation is that the spike in the number of cases that happened at that time was an isolated event unrelated to other activities on the island.
\begin{figure}[H]
    \centering
    \includegraphics[width=\linewidth]{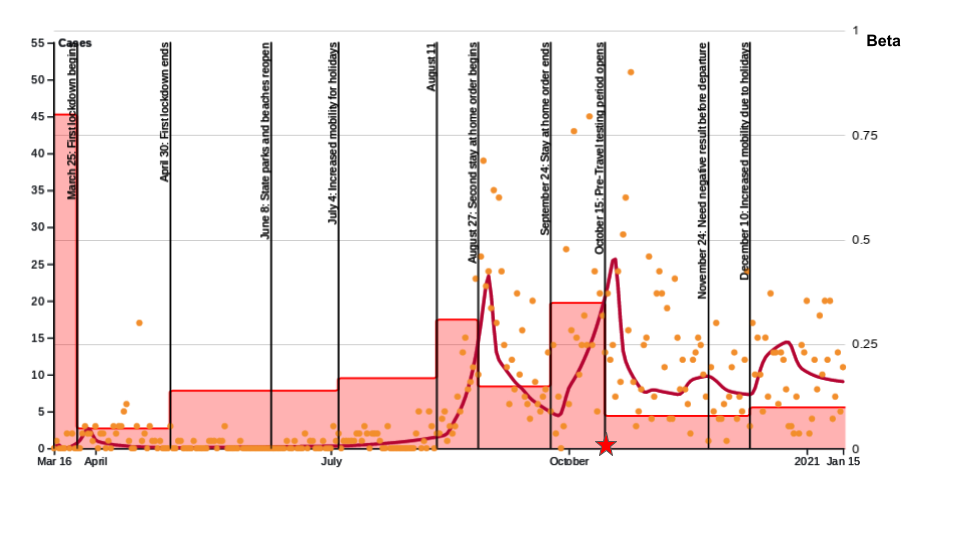}
    \caption{\Hawaii county: COVID-19 daily count and key events reflecting a change in behavior. The star shows the beginning of the safe travel program.}
    \label{fig:honolulu-plot}
\end{figure}

\begin{figure}[H]
    \centering
        \includegraphics[width=0.9\linewidth]{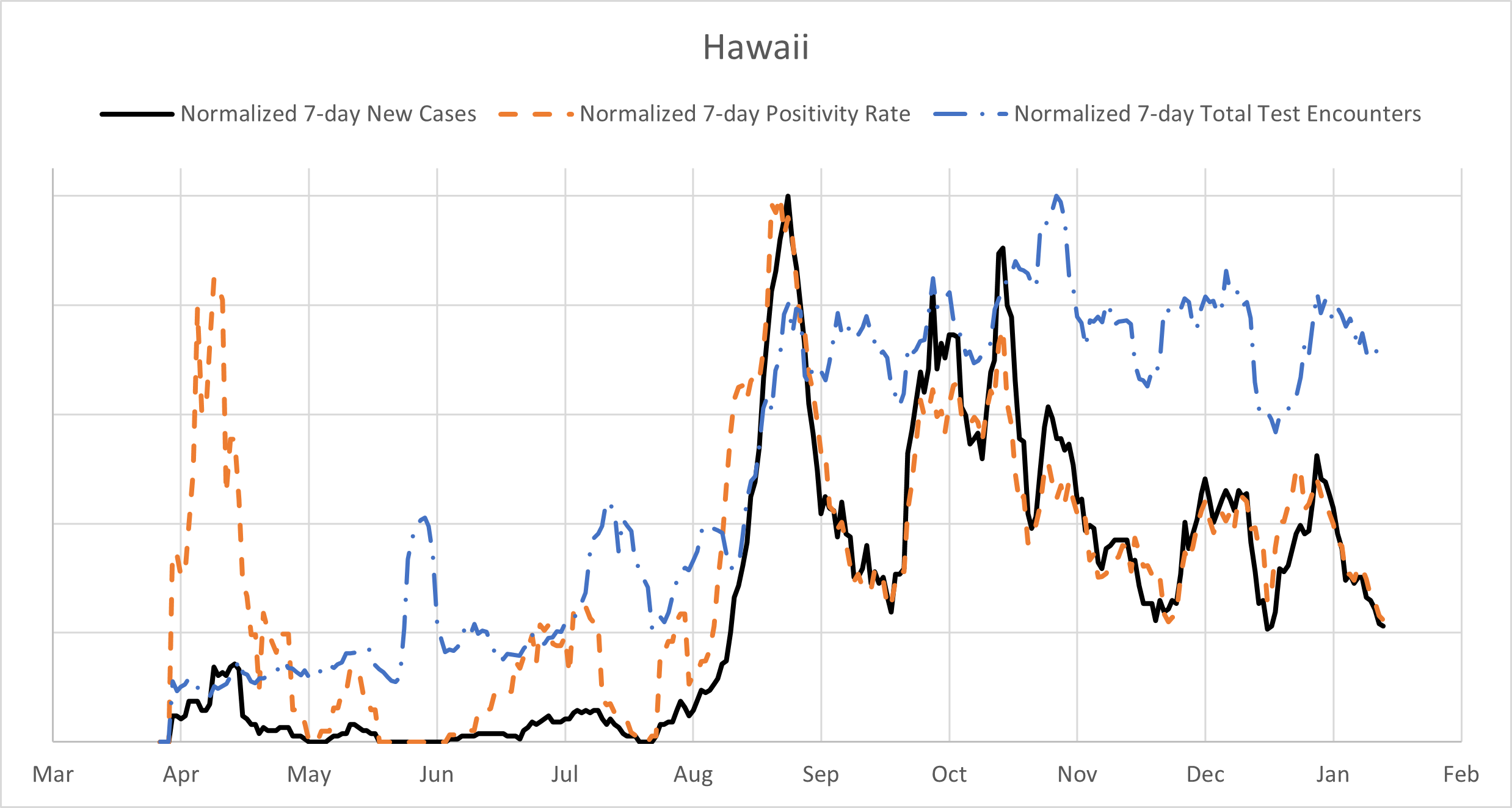}\\
    \includegraphics[width=0.9\linewidth]{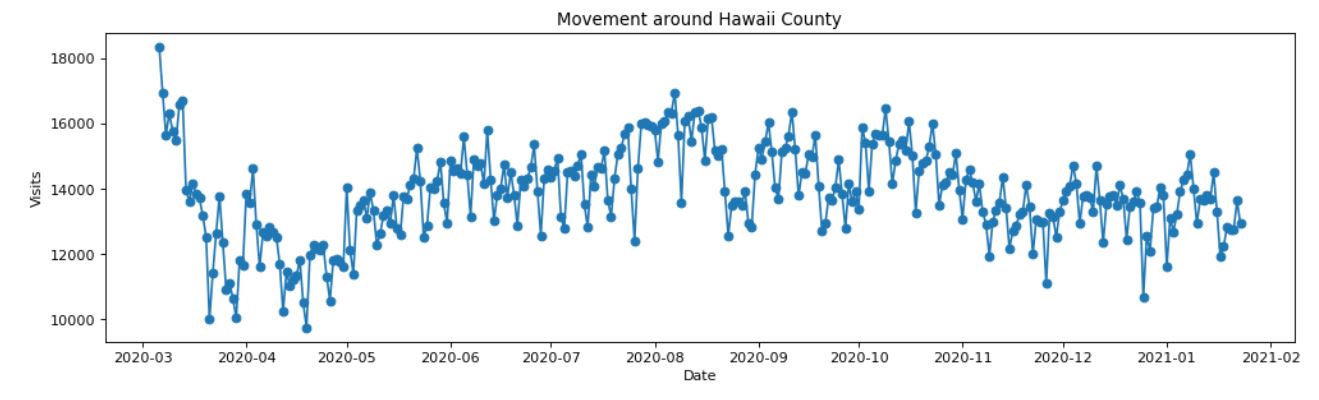}
    \caption{Top: A sharp increase in the test positivity rate around August indicates an outbreak the disease. The later decrease in the positivity rate with the number of tests hovering around the same value indicates a welcome slowdown of the spread of the disease. \ignore{In black solid: the 7-day rolling average of daily cases. In yellow dashed: the 7-day rolling average of positivity rate. In blue dashed-dotted: the 7-day rolling average of total tests. These data points are normalized to their maximums\cite{DOH}.} Bottom: Overall mobility for \Hawaii county from March 2020 to January 2021 indicates a mild correlation with the number of daily cases.}
    \label{fig:positivity-hawaii}
\end{figure}

\begin{figure}[H]
    \centering
    \includegraphics[width=0.6\linewidth]{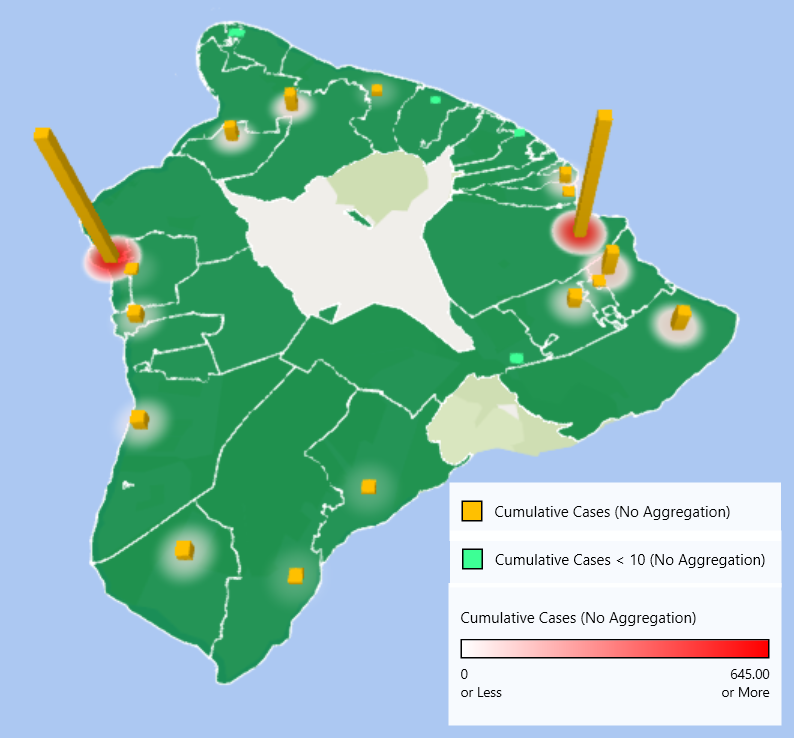}
    \caption{\Hawaii county cumulative daily counts distributed per zip code. }
    \label{fig:hawaii3D-zip-code}
\end{figure}

\begin{figure}[H]
    \centering
    \includegraphics[width=0.9\linewidth]{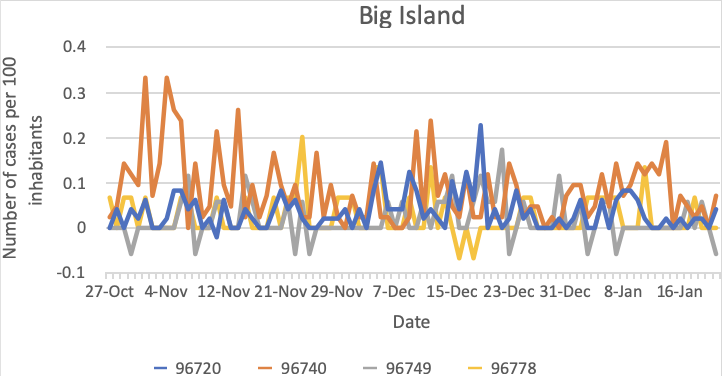}
    \caption{\Hawaii county cumulative daily counts distributed per zip code from October 2020 to January 18, 2021.}
    \label{fig:hawaii-zip-code-plots}
\end{figure}

\begin{table}[h!]
    \centering
    \begin{tabular}{ |p{3cm}|p{3cm}|p{3cm}|}
        \hline
        \multicolumn{3}{|c|}{Transmission rates} \\
        \hline
        Mar 16 - Mar 24 & Mar 25 - Apr 29 & Apr 30 - Jul 3\\
        \hline
        $\beta = 0.80$ & $\beta = 0.05$ & $\beta = 0.14$\\
        \hline
        Jul 4 - Aug 10 & Aug 11 - Aug 26 & Aug 27 - Sep 23\\
        \hline
        $\beta = 0.17$ & $\beta = 0.31$ & $\beta = 0.15$\\
        \hline
        Sep 24 - Oct 14 & Oct 15 - Dec 09 & Dec 10 - Jan 15\\
        \hline
        $\beta = 0.35$ & $\beta = 0.08$ & $\beta = 0.10$\\
        \hline

    \end{tabular}
    \caption{Optimized transmission rates to fit \Hawaii county data. They reflect the State and \Hawaii non-pharmaceutical mitigation measures. }
    \label{Table-transmissionrate-BigIsland}
\end{table}
On Fig.\ref{fig:positivity-hawaii} top we represent for the \Hawaii county the normalized total number of tests, the normalized tests positivity rate and normalized daily cases. Again test positivity correlates strongly with daily cases. The mobility for \Hawaii county did not show a decline as steep as for Honolulu county, and it shows good correlation with the daily cases and testing data.
Figure \ref{fig:hawaii3D-zip-code} shows the cumulative daily counts for \Hawaii county distributed per zip code from the onset of daily cases to January 18, 2021.  Clearly the vast majority of cases are located in one of the two main town: Kona (West) and Hilo (East).
Table \ref{tab:cumulative-data-zip-code-hawaii} summarizes the numbers for the four highest zip codes. Zip codes 96720 and 9674, respectively Hilo and Kona, clearly dominate the counts. We can see on Fig.\ref{fig:hawaii-zip-code-plots} that Kona had consistently larger number than Hilo for the exception of the few days before Christmas. This can be explained that overall the period October 15 to January 18, air traffic was quite more significant in Kona than in Hilo. Total (Tourist, Returning resident): Kona (76189,23824), Hilo (15808, 8800).
\begin{table}[h!]
    \centering
       \begin{tabular}{ |p{2.8cm}|p{2.8cm}|p{2.8cm}|p{2.8cm}|}
        \hline
        \Hawaii County Zip code & Population Estimate & Cumulative Daily cases & Cum. Daily cases per 100 inhabitants\\
        \hline\hline
         96720 & 48339 & 594 & 12 \\
         96740 & 42069 & 615 & 15 \\
         96749 & 17308 & 122 & 7 \\
         96778 & 14885 & 100 & 7 \\
        \hline
    \end{tabular}
    \caption{The four zip codes with the largest cumulative distribution of daily cases.}
    \label{tab:cumulative-data-zip-code-hawaii}
\end{table}
\subsection*{Maui County}
Maui county started the pandemic with a relatively low number of daily cases, but then entered an alarming state of a high number of cases per hundred thousand of population even reaching a  a maximum of 56 cases on January 6, 2021. It can be seen clearly the trigger with the introduction of the safe travel program on October 15, 2020. The influx of travelers is not constant through time and because the ratio tourists versus residents is high on Maui we see as a result the wavy increasing curve. In addition to the effect of additional tourists there was a large outbreak in relatively high population density condominium complex. The initial increase after October 15 was solely due to travelers which is why we see a rise in daily cases even though the basal transmission rate $\beta$ stays small.

\begin{figure}[H]
    \centering
    \includegraphics[width=\linewidth]{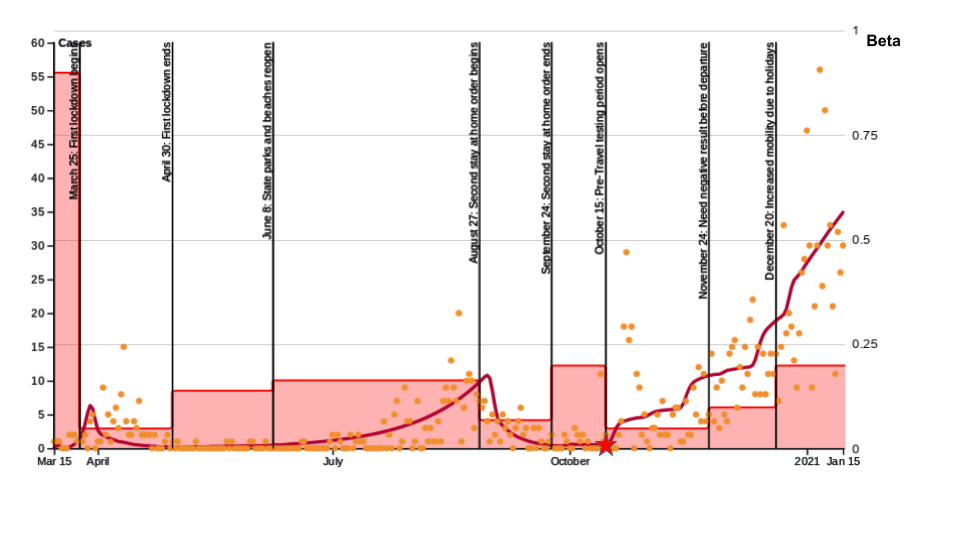}
    \caption{Maui county: COVID-19 daily count and key events reflecting a change in behavior. }
    \label{fig:maui-plot}
\end{figure}

\begin{figure}[H]
    \centering
        \includegraphics[width=0.9\linewidth]{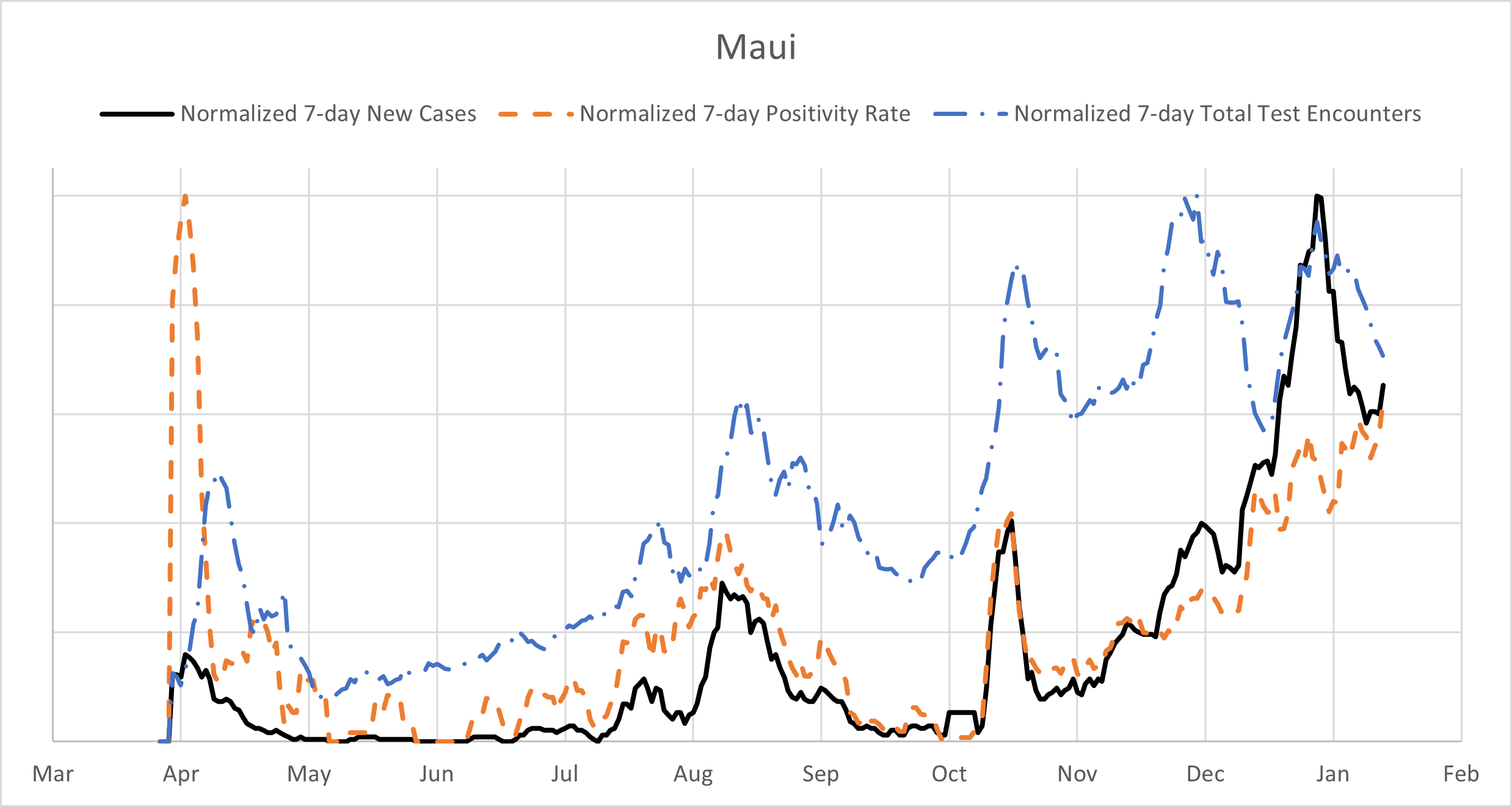}\\
    \includegraphics[width=0.9\linewidth]{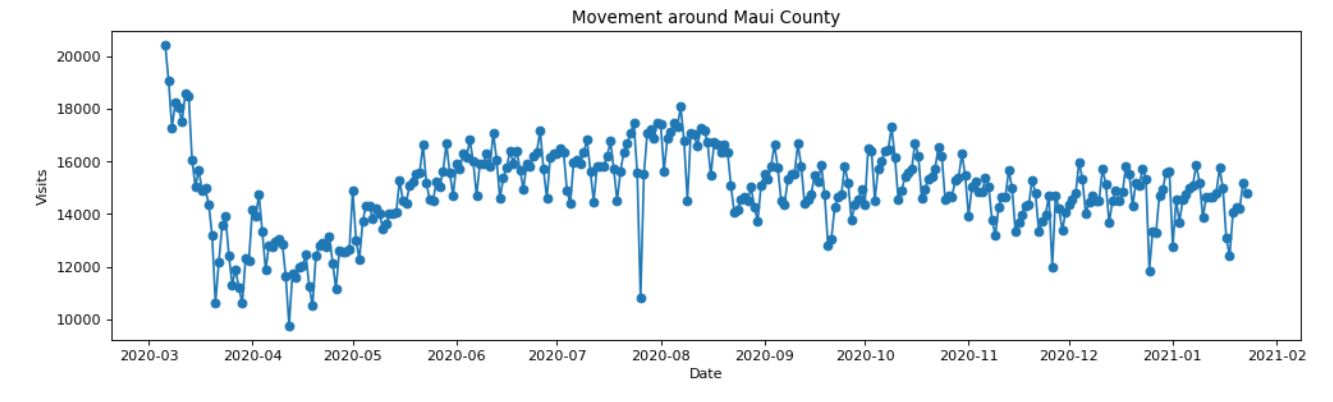}
    \caption{Top: A series of ups and downs in the test positivity rate and the number of daily cases indicate the occurrences of outbreaks of the disease. The significant increase in these numbers at the beginning of this year suggests a serious spread of the virus. A noticeable jump in the daily case number that does not correlate with the positivity rate can be explained by a jump in the number of tests, since the latter are performed for people with higher chances of having the virus. Bottom: Overall mobility for Maui County from March 2020 to January 2021 does not correlate well with the number of daily cases.}
    \label{fig:positivity-maui}
\end{figure}

\begin{figure}[H]
    \centering
    \includegraphics[width=0.7\linewidth]{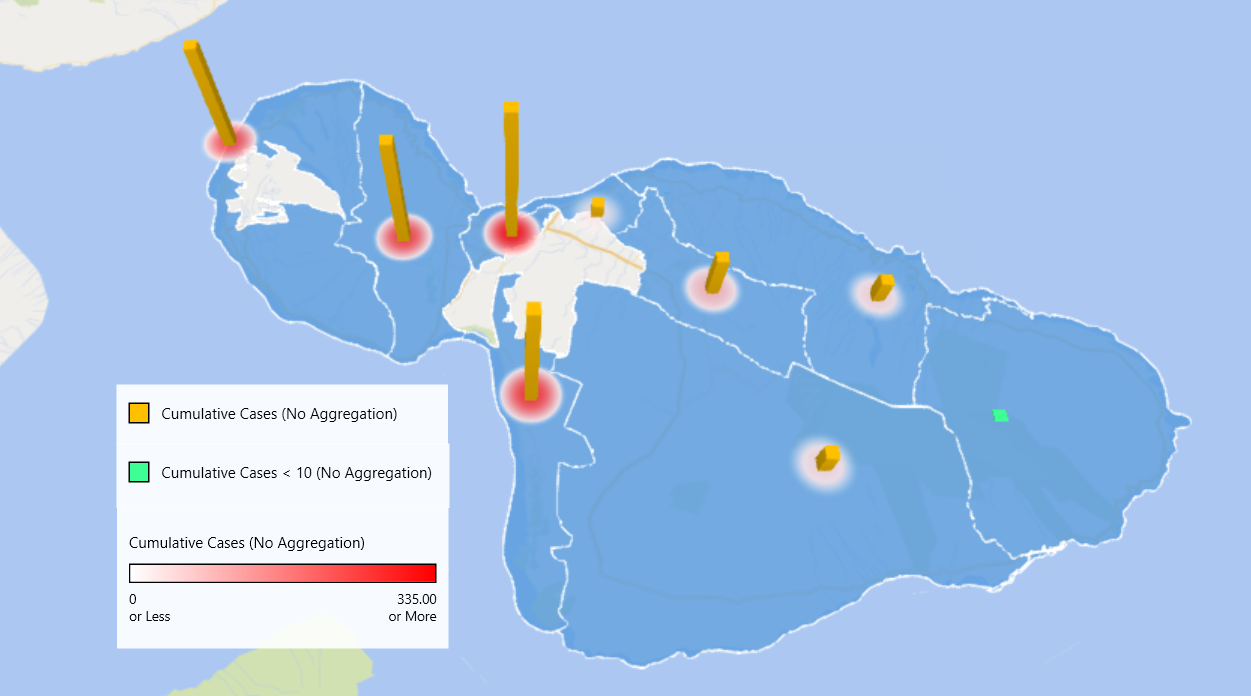}
    \caption{Maui county cumulative daily counts distributed per zip code. }
    \label{fig:maui3D-zip-code}
\end{figure}

\begin{figure}[H]
    \centering
    \includegraphics[width=0.9\linewidth]{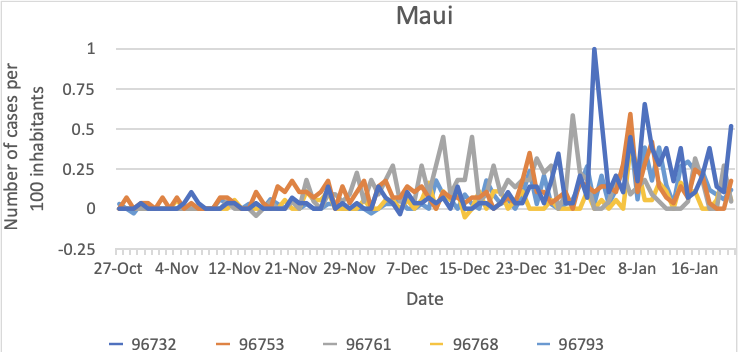}
    \caption{Maui county cumulative daily counts distributed per zip code from October 2020 to January 18, 2021. There were a few clusters on Maui which is the explanation for some of the higher spikes, in particular in early January in Kahului which is zip code 96732.}
    \label{fig:maui-zip-code-plots}
\end{figure}

\begin{table}[h!]
    \centering
    \begin{tabular}{ |p{3cm}|p{3cm}|p{3cm}|}
        \hline
        \multicolumn{3}{|c|}{Transmission rates} \\
        \hline
        Mar 15 - Mar 24 & Mar 25 - Apr 29 & Apr 30 - Jun 7\\
        \hline
        $\beta = 0.90$ & $\beta = 0.05$ & $\beta = 0.14$\\
        \hline
        Jun 8 - Aug 26 & Aug 27 - Sep 23 & Sep 24 - Oct 14\\
        \hline
        $\beta = 0.165$ & $\beta = 0.07$ & $\beta = 0.20$\\
        \hline
        Oct 15 - Nov 23 & Nov 24 - Dec 19 & Dec 20 - Jan 15  \\
        \hline
        $\beta = 0.05$ & $\beta = 0.10$ & $\beta = 0.2$\\
        \hline
    \end{tabular}
    \caption{Optimized transmission rates to fit Maui county data. They reflect the State and Maui non-pharmaceutical mitigation measures.}
    \label{Table-transmissionrate-Maui}
\end{table}
Tests, positivity and daily cases are represented on Fig.\ref{fig:positivity-maui} and show a strong correlation between the three. The mobility for Maui County seems correlating well until the introduction of the safe travel program. 
Figure \ref{fig:maui3D-zip-code} shows the cumulative daily counts for Maui county distributed per zip code from the onset of daily cases to January 18, 2021. The low counts on the eastern half of Maui are associated with low population density of local residents and relatively few tourists.
The four zip codes with the largest counts can be found in  \ref{tab:cumulative-data-zip-code-maui} and their daily behavior is displayed in Fig.\ref{fig:maui-zip-code-plots}. There was a large outbreak in a multistory  in early 2021 located in zip code 96732. The residents in this complex used elevators more than residents in other complexes in other areas with fewer stories. There are relatively larger number of tourists compared to local residents in zip codes 96761 and 96753 as compared to most other zip code areas. This is possible reason these two zip code area had larger increases in December than other areas. 
\begin{table}[h!]
    \centering
       \begin{tabular}{ |p{2.8cm}|p{2.8cm}|p{2.8cm}|p{2.8cm}|}
        \hline
        Maui County Zip code & Population Estimate & Cumulative Daily cases & Cum. Daily cases per 100 inhabitants\\
        \hline\hline
         96732 &  29075 & 278 & 9.5 \\
         96753 & 28737 & 259 & 9 \\
         96761 & 22301 & 240 & 11 \\
         96768 & 18529 & 90 & 5 \\
         96793 & 34036 & 211 & 6 \\
        \hline
    \end{tabular}
    \caption{The four zip codes with the largest cumulative distribution of daily cases.}
    \label{tab:cumulative-data-zip-code-maui}
\end{table}
\subsection*{\Kauai County}
\begin{figure}[H]
    \centering
    \includegraphics[width=0.8\linewidth]{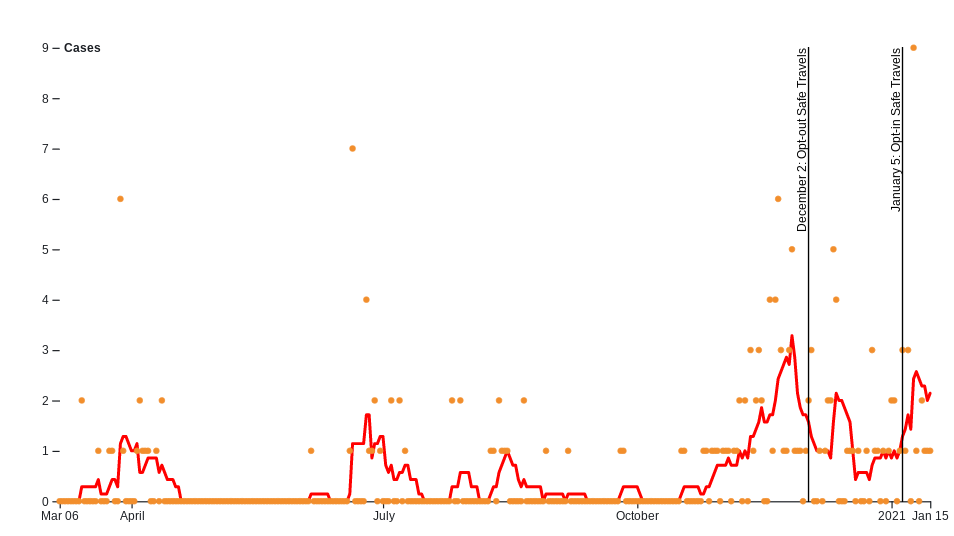}
    \caption{\Kauai county: COVID-19 daily count (orange) and 7-day average (red). The Opt-in Safe Travel on Jan 5 is only for intercounty travelers.}
    \label{fig:kauai-plot}
\end{figure}

\begin{figure}[H]
    \centering
        \includegraphics[width=0.9\linewidth]{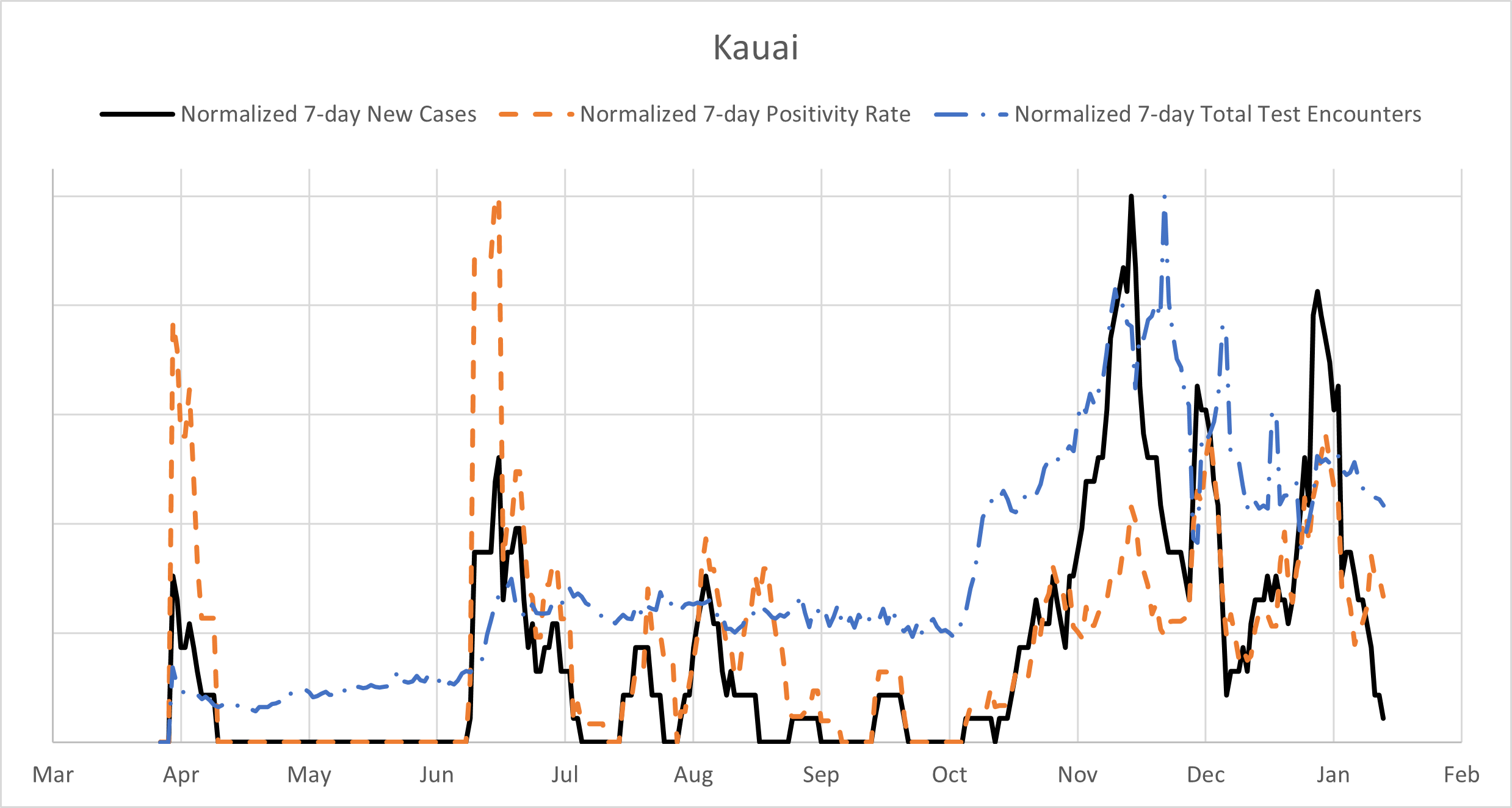}\\
    \includegraphics[width=0.9\linewidth]{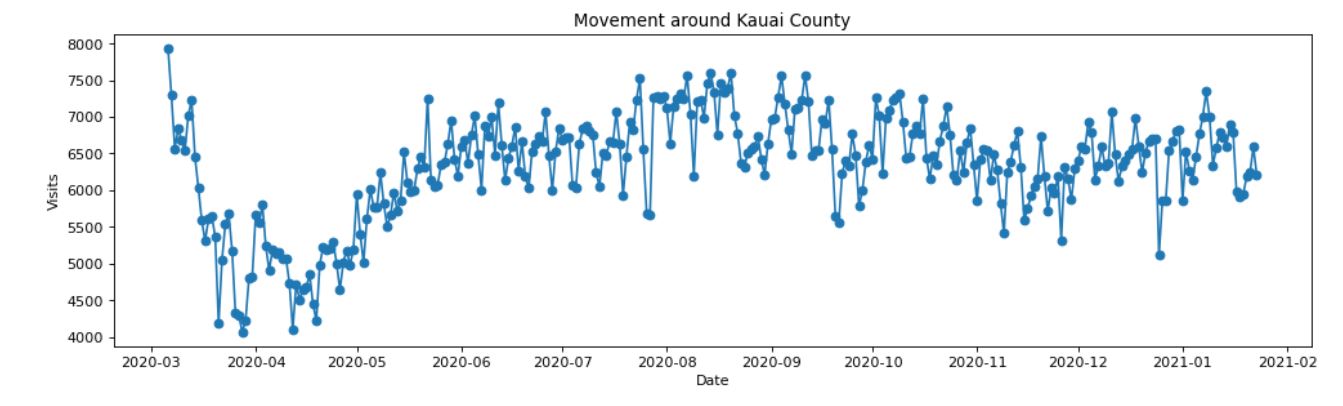}
    \caption{Top: The number of daily cases and test positivity rate are still well correlated, even though the raw numbers are small. Similar to \Hawaii county, we can see a jump in the daily case numbers that correlates with the increased number of tests rather than the test positivity rate, which is likely due to the biased nature of the population sample on which the tests are performed. Bottom: Overall mobility for \Kauai County from March 2020 to January 2021 does not correlate well with the daily cases.}
    \label{fig:positivity-kaui}
\end{figure}

Due to the low numbers on \Kauai a model fit using our compartmental model could not be achieved. It can be observed on Fig.\ref{fig:kauai-plot} that the daily cases started following an exponential growth, it was attributed to travelers which prompted the mayor of \Kauai to request authorization to opt-out from the safe travel program. It was followed by a decrease in numbers and stabilization. A new peak can be observed right after the safe travel program was authorized reinstated by \Kauai for intercounty travelers. The numbers are so small that is it extremely difficult to draw any additional conclusion. 
We still represent tests, positivity and daily cases on Fig.\ref{fig:positivity-kaui} and see as for Maui a strong correlation between the three. The mobility for \Kauai County is very flat after the initial decrease in March and even went a bit down after the safe travel program started. 
\section*{Discussion}
\begin{figure}[H]
    \centering
        \includegraphics[width=0.9\linewidth]{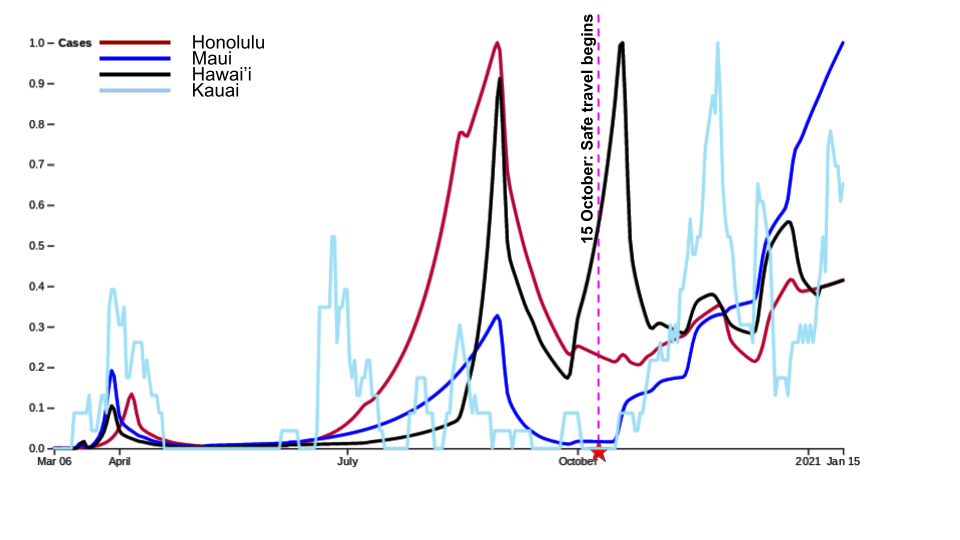}
    \caption{Honolulu, Maui and \Kauai counties with normalized model fit, \Kauai with normalized daily cases. It is clearly observed that counties started to differ in respond to the spread of COVID-19 after the safe travel program opened.}
    \label{fig:fit-all}
\end{figure}

There is clearly major differences among the four counties. Fig.\ref{fig:fit-all} shows on the same plot normalized model fits for Honolulu, \Hawaii and Maui counties as well as the daily raw numbers for \Kauai. It can observed that beside \Kauai for which numbers have been very low to draw comparison, the other three counties correlates well until when the Safe Travel program began on October 15, 2020. \Hawaii county displays an increase in daily numbers right before which were attributed to a couple of clusters (one in Hilo and one in Ocean View). Maui also had a few clusters, including a major one around October 20 on Lanai and another major one in early January in Kahului. After October 15, 2020 both Honolulu county and \Hawaii county show a slight increase in contrast with Maui county that displays a very sharp increase. Looking at Tables \ref{Table-transmissionrate-Oahu}, \ref{Table-transmissionrate-BigIsland} and \ref{Table-transmissionrate-Maui} we observe that the exponential growths and decays for \Hawaii and Maui counties require typically larger value for the basal transmission rate than for Honolulu county. The reason is that changes occurred more rapidly in the outer-islands, for instance the peak for Honolulu county is based on a build-up starting in June while for \Hawaii county the peaks are much more narrower. For Maui county the decay due to the second stay-at-home order was extremely efficient at the beginning and then slowed down which requested an increase in $\beta$. 
 We analyze similarity by using the $L_2$-norm for the difference between two normalized given model fits. One comparison was done over the entire length from March 6, 2020 to January 15, 2021. Results are displayed in Table \ref{table_hawaiian_county_L2}.  
\begin{table}[!h]
\centering
    \begin{tabular}{|c|c|c|}
        \hline
        &  Mar 06 - October 15 & Mar 06 - Jan 15\\
        \hline
        Honolulu - Maui  & 2.05353346 & 4.49831756\\
        \hline
        Honolulu - \Hawaii & 3.30655235 & 3.81229319 \\
        \hline
        \Hawaii - Maui & 3.53488371 & 4.23234035 \\
        \hline
    \end{tabular}
    \captionsetup{justification = centering}
    \caption {Normalized $L_{2}$ norm between hawaiian counties measuring similarity. Show the impact of the different county's regulations for travel since the dissimilarity between the counties grows when we add the period October 15, 2020 to January 15, 2021. In particular, before travel was instated Honolulu and Maui counties were the most similar, situation that reversed afterwards.}
    \label{table_hawaiian_county_L2}
\end{table}

 In regards to mobility and movement of the counties from the bottom halves of Fig. \ref{fig:positivity-honolulu}, \ref{fig:positivity-hawaii}, \ref{fig:positivity-maui}, and \ref{fig:positivity-kaui}, we acknowledge that the counties behavior follow a pattern that following the onset of the pandemic, movement dramatically slowed down. Afterwards, it began to plateau towards a movement index between the shutdown and normal. Curiously, the correlation between the mobility index and the daily case is far from strong, and in the case of \Kauai county the picture is more similar to anti-correlation (see Fig.\ref{fig:kauai-plot}). It suggests that the spread of the virus among households, especially large and multi-generational, could significantly contribute to the overall daily cases. As we can observe from the 3D zip code maps, the cases are very localized. Not surprisingly, they are higher in urban locations and towns where the population density, as well as the probability of indoor gatherings, is higher.

\begin{figure}[H]
    \centering
    \includegraphics[width=0.5\linewidth]{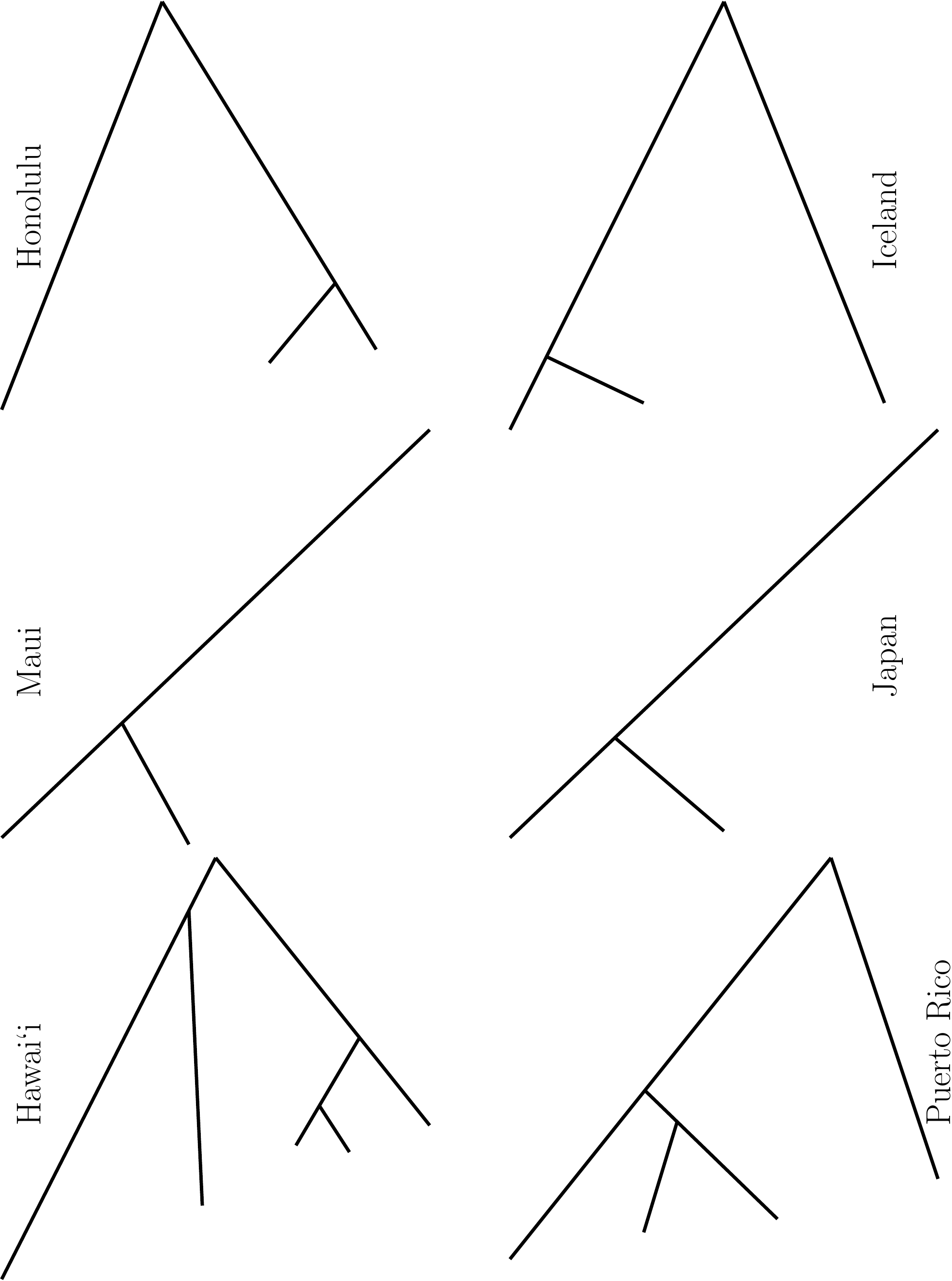}
    \caption{Merge trees elucidate the qualitative structure of the daily case numbers over time.}
    \label{fig:merge_trees}
\end{figure}

\begin{figure}[H]
    \centering
    \includegraphics[width=0.5\linewidth]{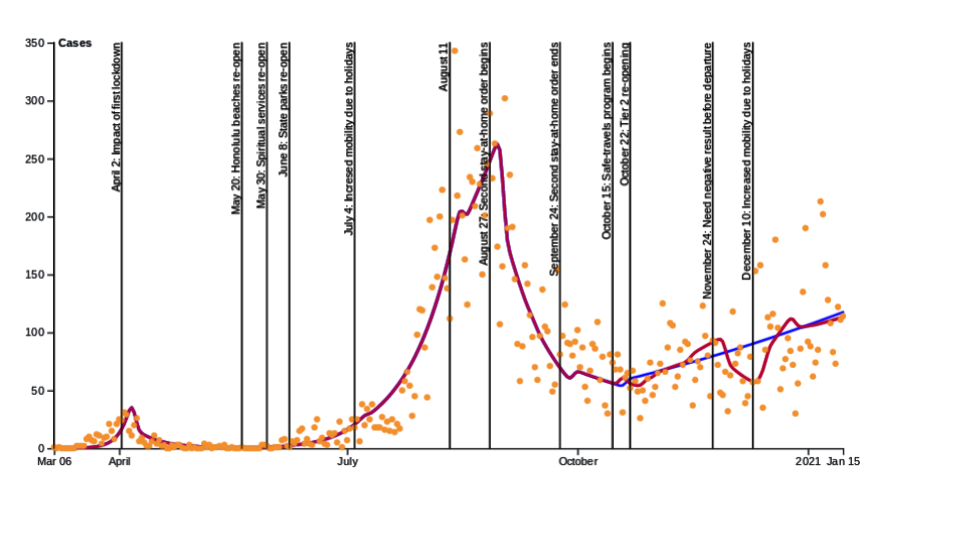}\includegraphics[width=0.5\linewidth]{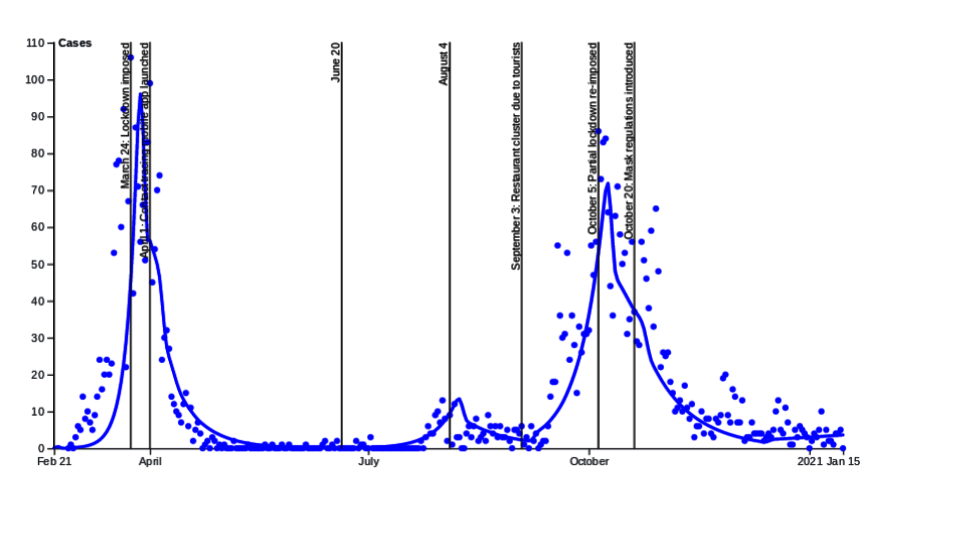}
      \includegraphics[width=0.5\linewidth]{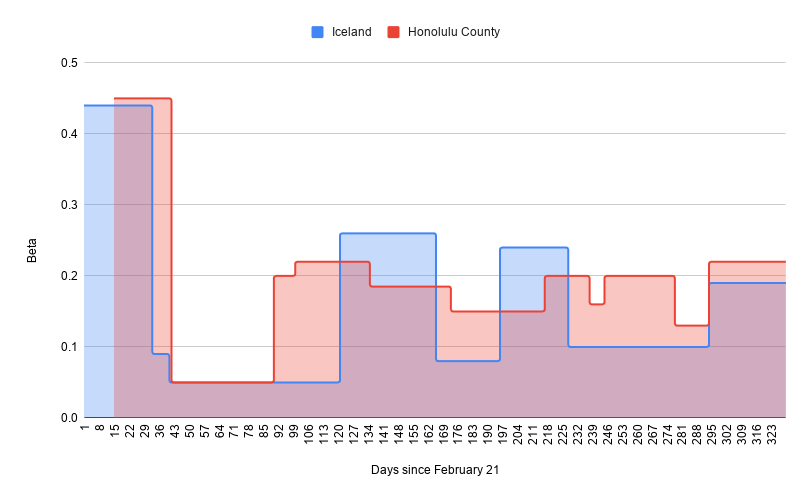}
    \caption{Top: Comparison between the daily cases between Honolulu County and Iceland. The blue fit for Honolulu correspond to not taking into account the travelers. Bottom: Superposition of the transmission rate values optimized for the fits above. }
    \label{fig:honolulu-iceland}
\end{figure}

\begin{figure}[H]
    \centering
    \includegraphics[width=0.5\linewidth]{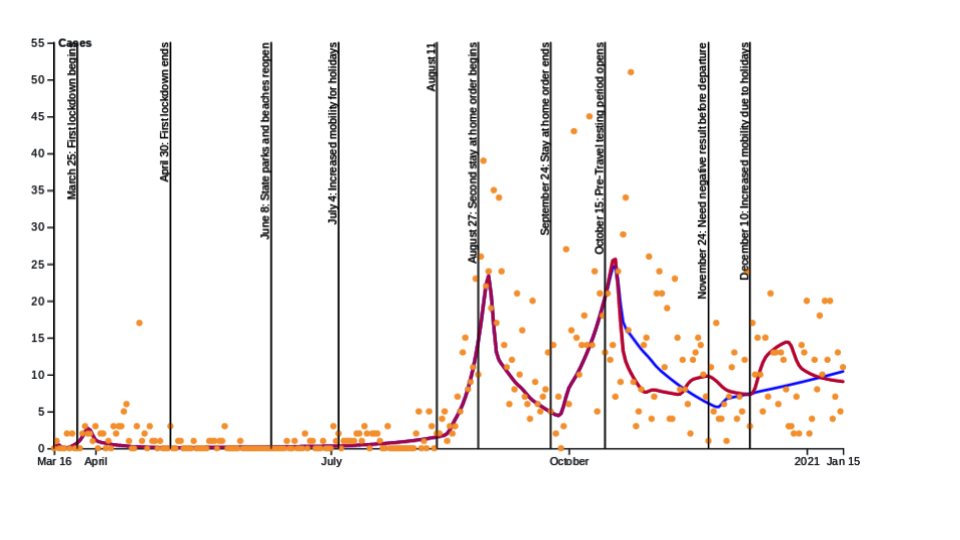}\includegraphics[width=0.5\linewidth]{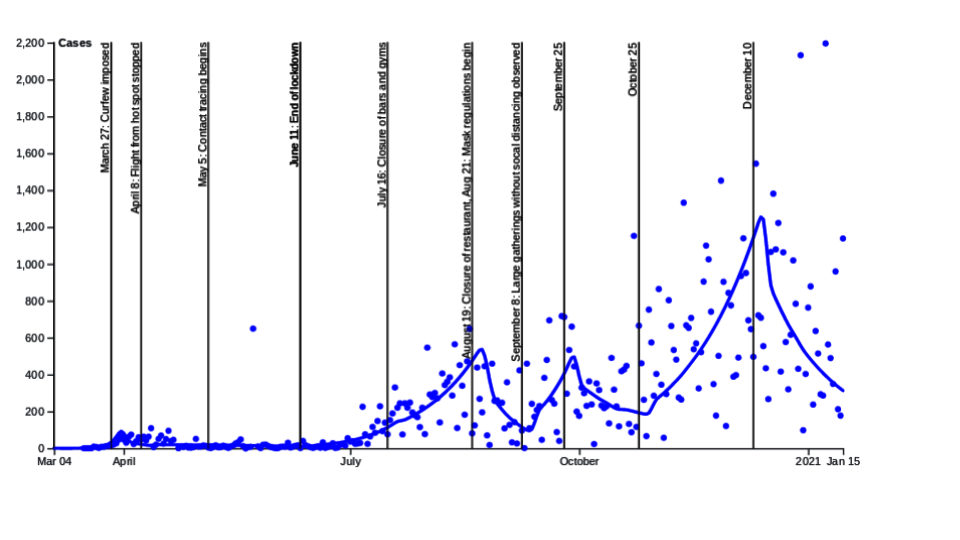}
     \includegraphics[width=0.5\linewidth]{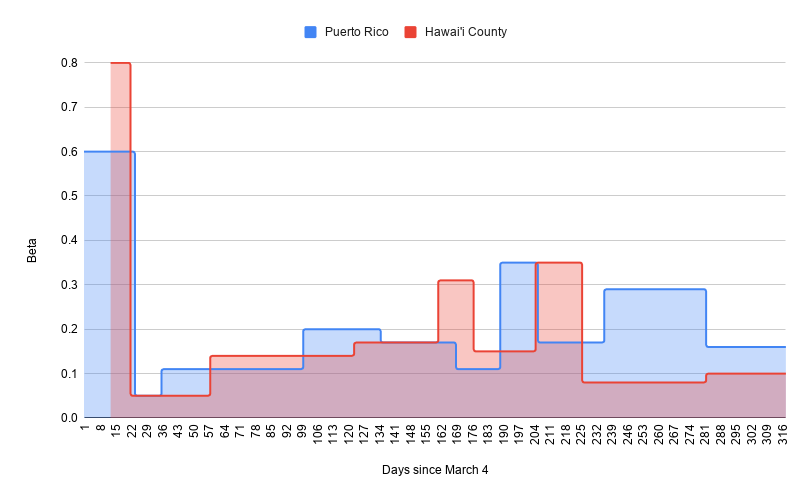}
    \caption{Top: Comparison between the daily cases between \Hawaii County and Puerto Rico. The blue fit for \Hawaii county is a smoothing occurring by neglecting the travelers. Bottom: Superposition of the transmission rate values optimized for the fits above.}
    \label{fig:PR-Hawaii}
\end{figure}

\begin{figure}[H]
    \centering
    \includegraphics[width=0.5\linewidth]{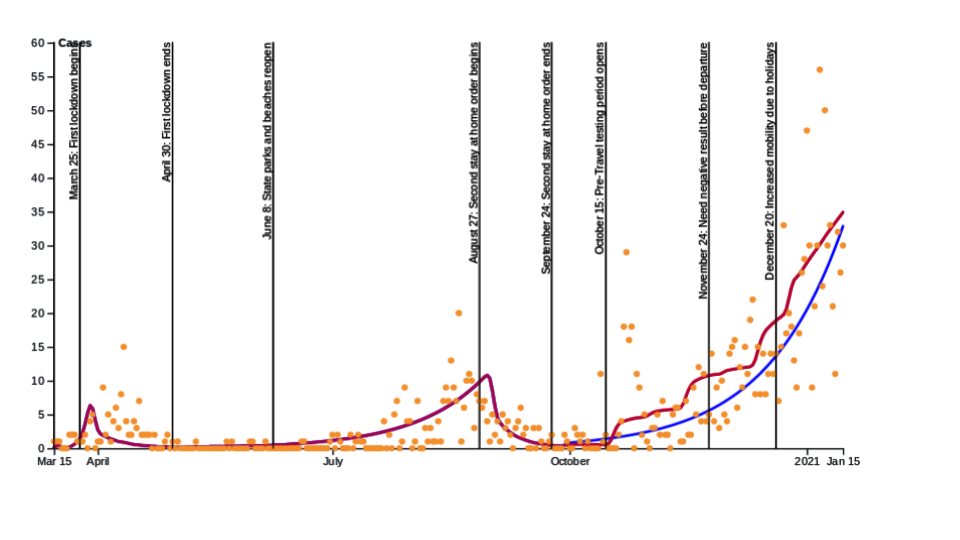}\includegraphics[width=0.5\linewidth]{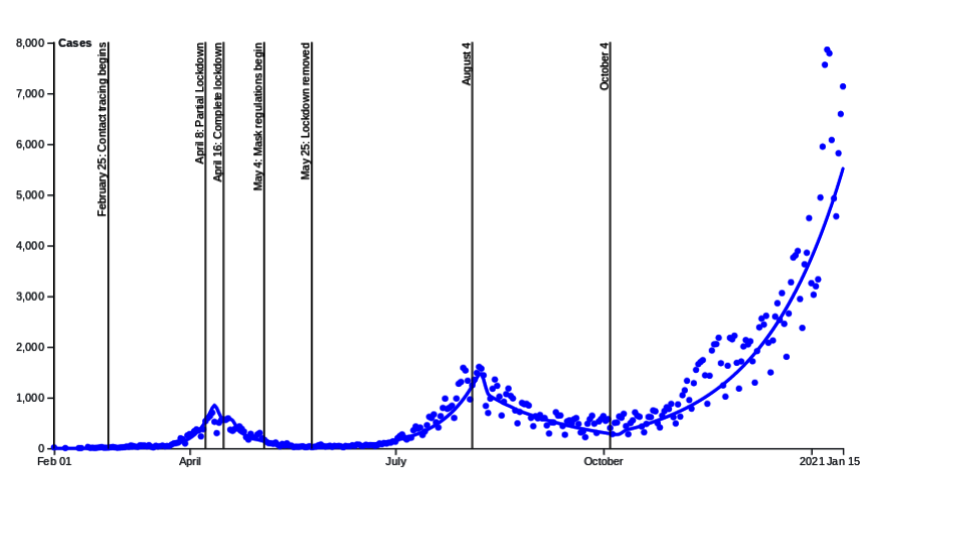}
     \includegraphics[width=0.5\linewidth]{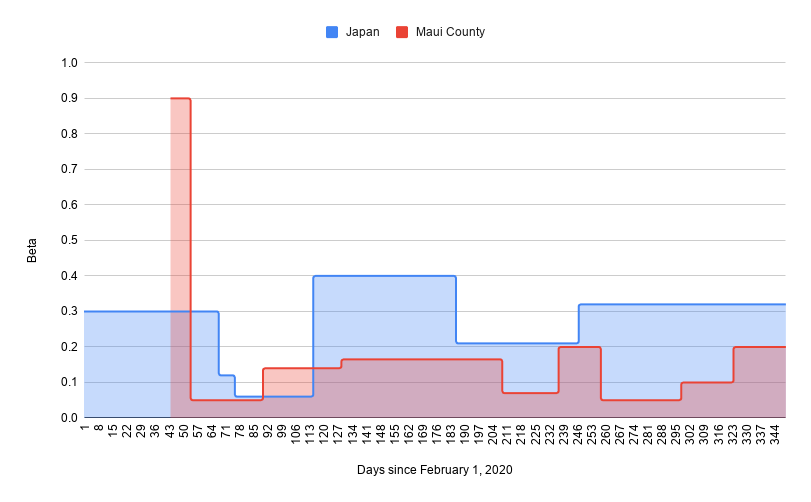}
    
    \caption{Top: Comparison between the daily cases between Maui County and Japan. Blue fit for Maui is without travelers. Bottom: Superposition of the transmission rate values optimized for the fits above.}
    \label{fig:Maui-japan}
\end{figure}

We use three other Islands: Japan, Iceland, Puerto-Rico for comparison with our counties. We ran a fit with our compartmental model for the three countries and analyze similarity by looking at the qualitative structure of the results as captured by \emph{merge trees} (see e.g. \cite{morozov_etal2013}). The latter construct is a topological descriptor of functions, and is constructed by tracking how connected components of the sublevel sets appear and merge as the threshold for the sublevel sets increases. This comparison is favored to a standard $L_{2}$ metric due to time shifts in the course of the pandemic for the various countries. An easy way to visualize this process is to move a horizontal line from the bottom to the top of the graph of a function and keep track of the function values at which a new connected component of the graph appears under the line or two existing components get merged. The actual horizontal locations of the branches, which represent the connected components, is not important, just their relative (left-right) positions. The merge trees for our three counties and the aforementioned countries are shown in Fig.\ref{fig:merge_trees}. They were computed using the normalized time series for the daily cases numbers starting from June 15. We also slightly simplified the structure (for illustration purposes) by removing very small branches. We can see that the Honolulu county merge tree is most similar to the Iceland merge tree and the Maui county merge tress is most similar to the Japan merge tree. The complexity of the \Hawaii county merge tree makes it more similar to the Puerto Rico merge tree, although these two are not as good of a match as the other pairs.
Figures \ref{fig:honolulu-iceland}, \ref{fig:PR-Hawaii} and \ref{fig:Maui-japan} display the model fit as well as the $\beta$ values for Honolulu, \Hawaii and Maui counties with their most similar Islands using the merge trees similarity of Fig.\ref{fig:merge_trees}. Table \ref{initial-data-countries} provides the initial values for Iceland, Japan and Puerto-Rico used by our model. travel restrictions vary widely between countries, see Table \ref{table-average-visitors} for the estimate made for Iceland, Japan and Puerto-Rico. 

Iceland, most similar to Honolulu County, detected their first case in February and had a significant first wave, but then controlled the spread beside a super spreader event trigger by two travelers. Traveling has then be very restricted which is why the daily cases are mostly in the single digits at the end of the fit. 
\Hawaii county is most similar to Puerto-Rico. The accuracy of the data for Puerto-Rico is unclear and it was very difficult to find the travel restrictions. The primary difference is the peak in December that Puerto-Rico suffered. 
Maui and Japan display a very similar qualitative curve, especially when travelers are ignored for Maui. The reason for Japan explosive growth at the end of the year is attributed to a few factors, including a controversial encouraging domestic travel policy that is as of January more restrictive and a possible COVID-19 fatigue by the population. 

\begin{table}[h!]
    \centering
    \begin{tabular}{|c|c|c|c|}
        \hline
         Region &  $S_c(0)$ & $S_h(0)$ & Date for $I_{c,0}(0)=1$\\
        \hline
        \hline
        Japan & 126500000 & 1673518 & Feb 01\\
        \hline
        Iceland & 356991 & 1404 & Feb 21\\
        \hline
        Puerto Rico & 3194000 & 89000 & Mar 04\\
        \hline
    \end{tabular}
    \captionsetup{justification=centering}
    \caption {Susceptible population for the three countries.}
    \label{initial-data-countries}
\end{table}

\section*{Conclusion}
In this paper, using the Hawaiian archipelago, we explore the importance of taking into account local variations in island chains. Ratios between residents and tourists as well as age demographic and other specificity call for targeted mitigation measures and Safe Travel program when in a pandemic. (Implementation of the Safe Travel program varies between the different Hawaiian counties.) The State of \Hawaii has launched an aggressive mass vaccination campaign starting in December but the effects of which are only now starting to impact the daily case rate. During the period of our study the very small impact of vaccination was neglected. As of February 2, 2021 we have 202,200 doses administered. The State policy is to keep the vaccination plan as originally planned to 2 doses per individuals even though two cases of the more transmissible B1.1.7 have been detected in \Hawaii. As of February 8, 2021 cases have been decreasing in all four counties. 

Not studied in this paper is hospital beds capacity, for the State of \Hawaii. The county of Honolulu is home to most of the hospital facilities and healthcare workers. The primary reason we did not discuss this here is the lack of consistent and clear data regarding hospitalisations. Similarly, quantification of COVID-19 related fatalities in the State of \Hawaii is delicate, indeed for instance in January about 60 deaths have been reclassified and added to the cumulative count. 

It is critical to conduct studies such as those presented here and capture critical data to be used in future pandemics. We are now working with the \Hawaii government on scenarios to understand impact of lifting some of the mitigation measures.

\section*{Acknowledgments}
This material is based upon work supported by the National Science Foundation under Grant No. 2030789.

\section*{Supporting information}
\paragraph*{S1 Fig.}
\label{S1_Fig-Zipcode}
{\bf Counties Zip Codes.} Zip Code Tabulation areas for the four counties. From the State of \Hawaii Office of Planning
2010 Census Reference Maps.
\begin{figure}[!h]
\includegraphics[width=0.5\linewidth]{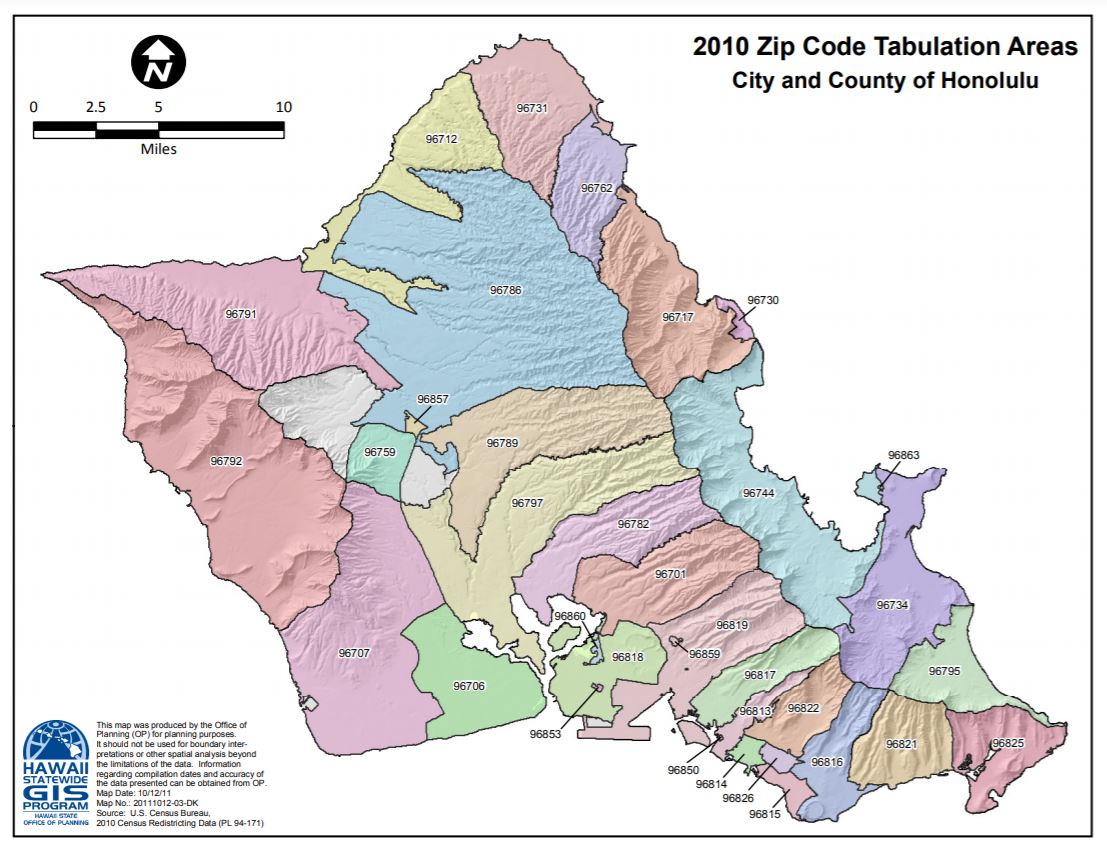}\includegraphics[width=0.5\linewidth]{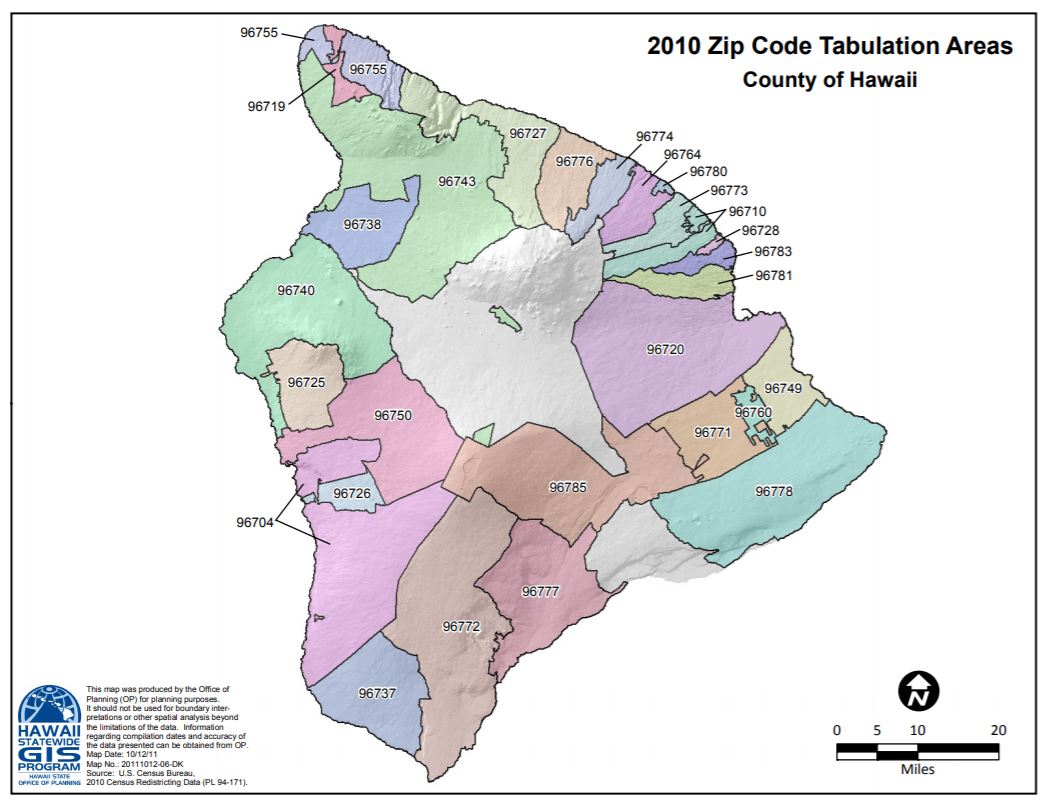}\\
\includegraphics[width=0.5\linewidth]{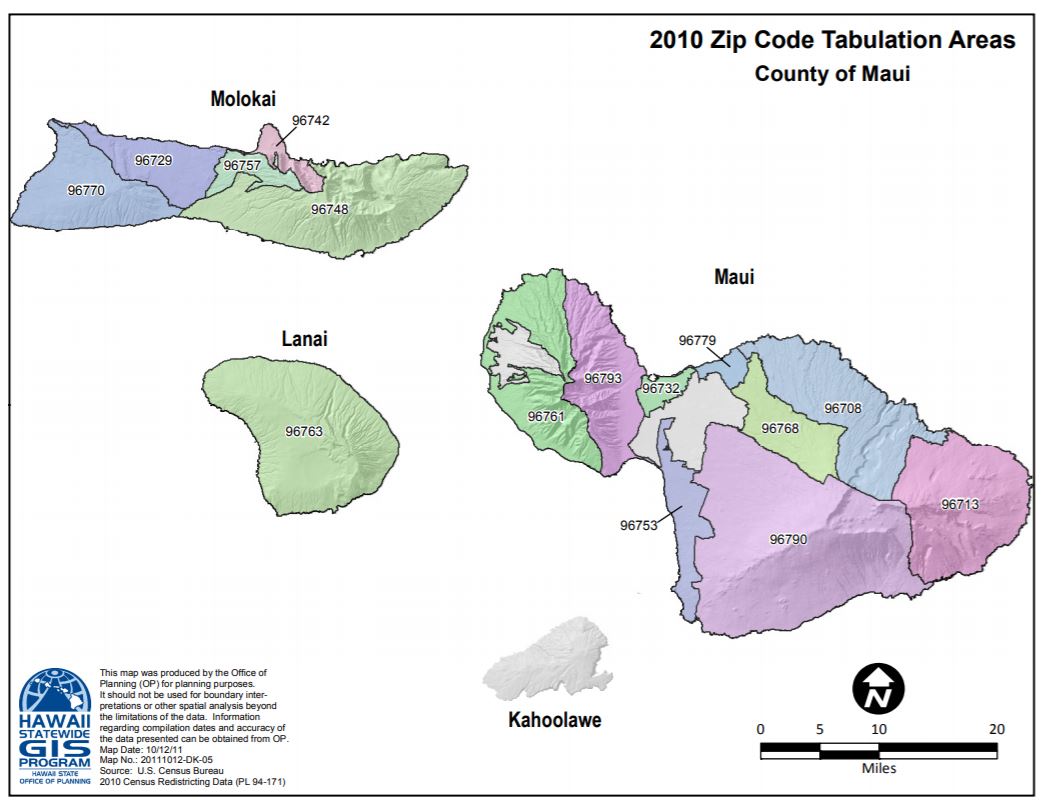}\includegraphics[width=0.5\linewidth]{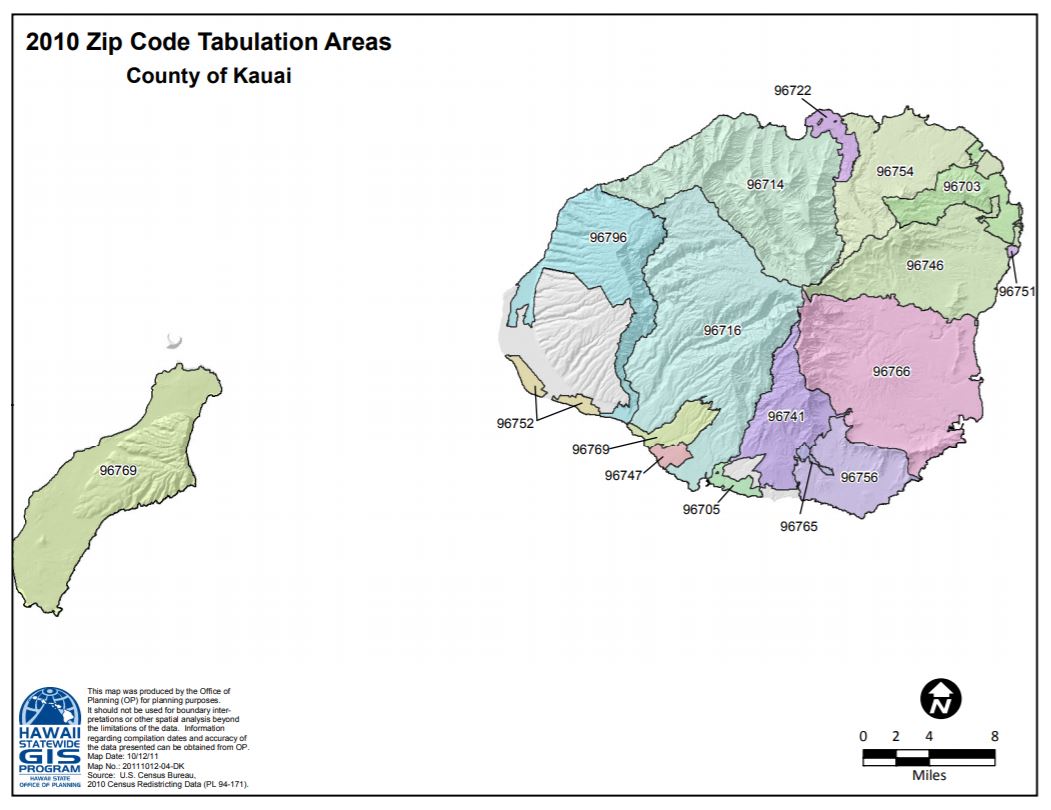}
\caption{{\bf ZipCodes per County.} Zip Code  Tabulation areas for the four counties. From the State of \Hawaii Office of Planning
2010 Census Reference Maps.}
\label{fig-zipcodes}
\end{figure}

\paragraph*{S1 Appendix.}
\label{S1_Appendix_Model}
{\bf Model Dynamics. }
The equations for the dynamics of the three population groups are essentially the same and are given below. Only the hazard rate and the parameters determining transition rates into quarantine may be different between the three groups.
\begin{align}
S(t+1) &= e^{-\lambda(t)}S(t)\\
E_0(t+1) &= (1-e^{-\lambda(t)})S(t)\\
E_{i}(t+1)& = (1-p_{i-1})(1-q_{a,i-1})E_{i-1}(t),\nonumber\\
&\quad\quad i=1,\ldots,13\\
E_{q,i}(t+1)& = (1-p_{i-1})(q_{a,i-1}E_{i-1}(t)+\nonumber\\
&\quad +E_{q,i-1}(t)), \quad i=1,\ldots,13\\
I_{0}(t+1) & =\sum_{i=0}^{13} p_i(1-q_{a,i})E_{i}(t)
\end{align}
\begin{align}
I_{1}(t+1) & =(1-q_{s,0})I_{0}(t)\\
I_{2}(t+1) & =(1-q_{s,1})I_{1}(t)+(1-r)(1-q_{s,2})I_{2}(t)\\
I_{j}(t+1) & =r(1-q_{s,j-1})I_{j-1}(t)+\nonumber\\
&\quad +(1-r)(1-q_{s,j})I_{j}(t),\quad j=3,4\\
I_{q,0}(t+1) & =\sum_{i=0}^{13} p_i(q_{a,i}E_{i}(t)+E_{q,i}(t))\\
I_{q,1}(t+1) & = I_{q,0}(t)+q_{s,0}I_{0}(t)\\
I_{q,2}(t+1) & = I_{q,1}(t)+q_{s,1}I_{1}(t)+\nonumber\\
&\quad +(1-r)(q_{s,2}I_{2}(t)+I_{q,2}(t))\\
I_{q,j}(t+1) & = r(q_{s,j-1}I_{j-1}(t)+I_{q, j-1}(t))+\nonumber\\
&\quad +(1-r)(q_{s,j}I_{j}(t)+I_{q,j}(t)),\quad j=3,4\\
R(t+1) &= R(t) + rI_{4}(t)+rI_{q,4}(t) + \nonumber\\
&\quad +(1-p_{13})E_{13}(t) + (1-p_{13})E_{q,13}(t)
\end{align}
Below is a detailed description of the variables, all of which depend on time, $t$, measured in days.
\begin{itemize}
 \item\textbf{Variable $S(t)$.} The number of susceptible individuals.
 \item\textbf{Variables $E_{i}(t)$.} The number of asymptomatic infected individuals $i$ days after exposure who are not quarantined.\\
 \item \textbf{Variables $E_{q,i}(t)$.} The number of quarantined asymptomatic infected individuals $i$ days after exposure.\\
 \item \textbf{Variables $I_{j}(t)$, $i=0,1$.} The number of symptomatic infected individuals $i$ days after the onset of symptoms who are not quarantined.\\
 \item \textbf{Variables $I_{j}(t)$, $j=3,4,5$.} The number of symptomatic infected individuals at the nominal stage $i$ of the illness. Note that a person can stay at a given stage for several days.\\
 \item \textbf{Variables $I_{q,j}(t)$, $j=0,1$.} The number of quarantined symptomatic infected individuals, with $j$ representing either the number of days after the onset of the symptoms ($j=0,1$), or the stage of the illness ($j=2,3,4$).\\
 \item \textbf{Variable $R(t)$.} The number of removed (recovered or deceased) individuals.
 \end{itemize}
Splitting exposed individuals into multiple stages, $E_{i}$, allows us to capture possible differences in the progression of the asymptomatic phase of the disease. Importantly, it allows us to take into account that, according to the Centers for Disease Control and Prevention (CDC) as well as other sources, about 40\% of people who contract SARS-CoV-2 remain asymptomatic, and the incubation period for those who do develop symptoms is somewhere between 2 to 14 days after exposure, with the mean incubation period between 4 and 6 days \cite{park2020systematic,aoim,pnas}. Individuals who do not develop symptoms after 14 days are assumed recovered. The use of the quarantine sub-compartments, $E_{q,i}$, allows us to capture the effect of contact tracing and the reduced transmission rate for quarantined individuals.

Similarly, having multiple stages for infected individuals better reflects progression of the symptomatic phase of the disease. The first two stages represent the first two days of symptoms, but the next three should be understood as phases of the immune system fighting the disease. There is a substantial variability (due to age as well as other factors) in the number of days any given person can spend at each stage. Our model implicitly assumes that the symptomatic phase of the illness lasts at least 5 days (in the unlikely case that each stage lasts just one day). 
 
 As we mentioned, a crucial part of the dynamics relates to the hazard rate. For the general community, group C, we have
\begin{multline}
\lambda_c(t) = \beta(1-p_{mp}(1-p_{me}))\Big[
  (I_c+\varepsilon E_c)+
  \gamma((1-\nu)I_{c,q}+\varepsilon E_{c,q})+\\
  \rho[(I_h+\varepsilon E_h)+  \gamma((1-\nu)I_{h,q}+\varepsilon E_{h,q})]+\\
   \rho_{v}[(I_v+\varepsilon E_v)+  \gamma((1-\nu)I_{v,q}+\varepsilon E_{v,q})]\Big]/(N_c+\rho_{v}N_v),
\end{multline}
 and for the tourists we have
\begin{multline}
\lambda_v(t) =\frac{ \rho_{v} \beta\lambda_c+ 
  \beta_v(1-p_{mp}(1-p_{me}))\Big[
  (I_v+\varepsilon E_v)+
  \gamma((1-\nu)I_{v,q}+\varepsilon E_{v,q})\Big]}{(\rho_{v}N_c+N_v)},
\end{multline}
where we suppressed the dependency on $t$ on the right for convenience. We use sub-indices $c$ (community), $h$ (healthcare workers), and $v$ (tourists) to indicate the appropriate group. Subscript $q$ indicates quarantined individuals. Here $p_{me}$ and $p_{mp}$ represent mask efficiency and mask compliance. Mask efficiency is chosen to reflect a reduction in transmission of $75\%$ for all regions. Mask compliance is set at $20\%$ for all regions at the start of the pandemic, but this value is modified on the dates the regions introduce mask regulations. $N_{v}$ denotes the mixing pool for the visitors and $N_c$ denotes the mixing pool for the general community, computed as
\begin{equation}
    N_c(t) = S_c+E_c+I_c+R_c+\rho(S_h+E_h+I_h+R_h)+\rho_{v1}(S_v+E_v+I_v+R_v).
\end{equation}
where variables $E$ and $I$ here represent the sum over all the stages within these compartments. 
For the healthcare worker group, we have
\begin{equation}
    \lambda_h(t) = \rho\lambda_c+\beta\eta\Big[
  (I_h+\varepsilon E_h)+
  \kappa\nu(I_{h,q}+I_{c,q}+I_{v,q})\Big]/N_h,
\end{equation}
where $N_h(t) = S_h+E_h+I_h+R_h$. 

The model fit plot the following value
\begin{equation}
    \label{eq-fitplot}
    \sum_{x=c,h,v}\Big(\sum_{i=1}^3 q_{s,i}^xI(i)+(1-r) q_{s,4}^xI(4)+\sum_{i=1}^12 q_{a,i}^xE(i) \Big)
\end{equation}

The safe travel program started on October 15, 2020 which is when travelers are implemented in the model (they were negligible before that). Based on Safe Travels Digital Platform from the State of Hawai'i, we are assuming a pre-travel testing rate of $86\%$, a false negative rate of $0.5\%$. We also assume $1\%$ of untested visitors go into exposed quarantine (we had to remove exempt travellers) and a $5\%$ prevalence for the virus. The pre-testing rates for travelers to Maui county is higher, and assumed  to be $95\%$. Traveler average influx is modeled as a piece-wise linear function  over two week intervals between the aforementioned time interval. The average influx for the comparing countries is assumed for simplification to be linear over the same time period.  See Table  \ref{table-average-visitors}.   
\begin{table}[h!]
    \centering
    \begin{tabular}{|c|c|c|}
        \hline
         Region &  Tourists & Returning Residents\\
        \hline
        \hline
        Honolulu & [1353,2124,3051,2028,4724,2195]  & [692,716,967,951,1014,1018]  \\
        \hline
        Maui & [800,1000,2000,1700,3000,2500] & [128,127,135,158,160,156]  \\
        \hline
        \Hawaii & [297,593,981,751,1712,1000]& [116,113,108,136,124,128]\\
        \hline
        Japan &700  &700  \\
        \hline
        Iceland & 0 & 0  \\
        \hline
        Puerto Rico & 3500 & 500  \\
        \hline
    \end{tabular}
    \captionsetup{justification=centering}
    \caption {Average visitors per day, starting on October 15, 2020 to January 15, 2021.}
    \label{table-average-visitors}
\end{table}

\end{document}